\documentclass[12pt]{article}

\usepackage{graphicx}
\usepackage{epsfig}
\usepackage{psfrag}
\usepackage{wrapfig}
\usepackage[all]{xy}
\usepackage{bbm}

\usepackage{fullpage}
\usepackage{setspace}
\usepackage{enumerate}
\usepackage{url}
\usepackage{xcolor}
\usepackage{ctable}
\usepackage{dcolumn}
\usepackage{multirow}

\usepackage{amsmath}
\usepackage{amssymb}
\usepackage{amsthm}
\usepackage{url}

\usepackage{footmisc}
\newcommand\wordcount{
    \immediate\write18{texcount -sum -1 \jobname.tex > 'count.txt'}
\input{count.txt}words}

\usepackage{natbib}

\usepackage{amsmath}
\usepackage{amsfonts}
\usepackage{amsthm}
\usepackage{bbm}
\usepackage{array}

\newtheorem{prop}{Proposition}

\newtheorem{assn}{Assumption}

\newcommand{\bi}{\begin{itemize}}
\newcommand{\ei}{\end{itemize}}

\def\Var{{\rm Var}\,}
\def\E{{\rm E}\,}

\def\Cov{{\rm Cov}\,}

\def\If{\mathbb{I}\,}

\newcommand\independent{\protect\mathpalette{\protect\independenT}{\perp}}
\def\independenT#1#2{\mathrel{\rlap{$#1#2$}\mkern2mu{#1#2}}}

\begin{document}

\sloppy
\author{Dana Burde, Joel Middleton, Cyrus Samii and Ye Wang\thanks{Dana Burde is Associate Professor and Director of International Education at the Steinhardt School of Culture, Education, and Human Development, New York University (Email: dana.burde@nyu.edu).  Joel Middleton is Assistant Professor of Political Science at the University of California, Berkeley (email: joel.middleton@berkeley.edu). Cyrus Samii (contact author) is Associate Professor of Politics at New York University (Email: cds2083@nyu.edu). Ye Wang is PhD Candidate in Politics at New York University and Doctoral Fellow at University of California, San Diego (Email: yw1576@nyu.edu).  This research comes as part of the {\it Assessment of Learning and Social Effects of Community Based Education in Afghanistan} (ALSE) project (www.alseproject.com), supported by a grant from the United States Agency for International Development. It is approved by the New York University institutional review board.  Special thanks to the ALSE research and management team, Amy Kapit, Jo Kelcey, Matthew Lisiecki, Vedrana Misic, Otgonjargal Okhidoi, Mohammad Amin Sadiqi, Cornelia Sage, and Susana Toro Isaza, and to the ALSE project associates Zabihullah Buda, Hamidullah Gharibzada, Mustafa Hasani, Abdul Hamid Hatsaandh, Ahmad Tamim Naseh, Mohammad Jawed Nazari, Iqbal Ali Shahrwand, and Mohammad Amin Zafari.  For helpful comments, we thank seminar participants at AALIMS, the Center for Global Development, and the Oxford University Global Priorities Institute.}}

\title{How to Account for Alternatives When Comparing Effects: Revisiting `Bringing Education to Afghan Girls'}
\maketitle

\pagenumbering{gobble}

\clearpage

\setcounter{page}{0}
\begin{center}
{\LARGE How to Account for Alternatives When Comparing\\
\medskip
 Effects: Revisiting `Bringing Education to Afghan Girls'}
\end{center}

\vspace{1in}

\begin{abstract} 
This paper uses a ``principal strata'' approach to decompose treatment effects and interpret why a schooling intervention that yielded exceptional initial effects yielded substantially smaller effects in a replication years later.
The specific application is a set of 2008 and 2015 replications of an intervention aiming to increase primary education for girls in rural Afghanistan.
The intervention offers a new schooling option, and as such, its effects depend on how individuals use alternatives that already exist.
The principal strata approach accounts variation in use patterns when comparing effects across the replications.
Our findings show that even though the share of girls for whom the intervention would be valuable dropped considerably in 2015 as compared to 2008, the intervention was even more efficaciousness for those who continued to benefit from it.
\end{abstract}

\centerline{Version 1.0}

%
\clearpage
\pagenumbering{arabic}
\doublespace

\section{Introduction}

This paper uses a ``principal strata'' approach to decompose treatment effects and interpret why a replication of a highly successful schooling intervention studied by \citet{burde2013bringing} yielded subsequent average effects that were substantially smaller. The intervention seeks to increase primary education for girls in rural Afghanistan, a context where access to education has been impeded by years of war and low state capacity \citep{rashid2008descent, burde2014schools}.
Reducing gender inequalities and development prospects more generally depends crucially on improving girls' access to education \citep{jayachandran2015roots, sperling2015works, muralidharan2017cycling}. 
This in turn makes it important to understand factors that affect the outcomes of attempts to improve access.
Replications over time and across contexts are crucial for informing our understanding of such factors \citep{rodrik2008new, banerjee2009experimental, vivalt2020much, rosenzweig2020external, dehejia2021local}. 
In this paper, we go beyond just reporting on variation in average effects across replications. 
We try to interpret such variation in effects in terms of the changes in the schooling options available \citep{kowalski2016doing}. 
This allows us to evaluate for what share of the population, in what ways, and to what extent the intervention might still be valuable.

Our analysis tackles a general problem of comparing effects of an intervention across situations that vary in terms of subjects' use of existing alternatives to the treatment.  
Principal strata refer to subpopulations that differ in terms of potential outcomes, such as choices that would be made in the absence of treatment \citep{frangakis_rubin2002, page2015principal, feller2017principal}.
For example, when one intervenes in a population, the effects may depend on whether intended beneficiaries find that the opportunities afforded by the intervention are appealing enough to switch over from using an existing alternative.  
In the application in this paper, a new form of school is introduced into Afghan villages.  
It stands to reason that the effect would depend on which families use the new form of school, which stick with using existing schools, and which continue not to attend school at all.  
Use patterns depend on how subjects assess opportunities that treatments afford relative to available alternatives \citep{chassang-etal-selective}.  
Such differences define the principal strata.
When the same intervention is applied in another context (e.g., years later, as in this paper, or in another population), use patterns can differ, in which case the composition of the principal strata may differ, as may the effects of the intervention.  
Our analysis demonstrates ways to derive insight from a set of experimental results in which use of alternatives varies.

It is now standard to consider compliance versus non-compliance with treatment assignments when analyzing experiments. 
One can estimate the local average treatment effect (LATE) for the compliers by using an instrumental variables analysis \citep{angrist_etal96_late}.  
Sometimes this set-up is too sparse, however.  
In our schooling example, there was both a pre-existing form of school and the option for children not to go to school at all.  
\citet{frangakis_rubin2002} introduced the term ``principal strata'' for a partition of the population on the basis of potential responses to a treatment. 
In the LATE analysis, we have the well known stratification into ``always takers,'' ``never takers,'' ``compliers,'' and ``defiers.''  
When we think of choices that go beyond just compliance versus non-compliance, the number of principal strata increases, up to the square of the number of choices.   
The partial identification approach in this paper provides bounds for effects over a richer set of effects over the principal strata (i.e. ``principal effects'').  
As we will show below, this will allow us to disentangle potential gains for those who had previously not used any school from potential harms for those drawn away from using the pre-existing school alternative.  
Partial identification along the lines considered here is sufficient for deriving optimal policy choices under criteria such as minimax regret \citep{manski2009identification}.

Our approach builds on some recent work. 
Our partial identification approach and the derivations of our inferential results build directly on \citet{lee2009training}, who derived non-parametric trimming bounds on treatment effects under attrition or sample selection.  
A few recent papers pursue substantive goals very similar to ours. 
\citet{kline_walters2016}, \citet{feller2016compared}, and \citet{kamat2020identification} study the Head Start pre-school program in the United States, where the potential for substitution between Head Start and other close substitutes makes the situation structurally identical to ours. 
\citet{kline_walters2016} and \citet{feller2016compared} apply the same monotonicity and exclusion assumptions that we state below and derive the same types of principal strata as we do. 
\citet{kline_walters2016} then use an economic model of school choice to motivate an interpretation of the LATE and take advantage of cross-site variation in take-up patterns to point identify principal effects under the restrictions of the economic model. 
\citet{feller2016compared} use a Bayesian normal finite mixture model to identify the principal effects. 
\citet{kamat2020identification} uses a non-parametric choice model to identify and partially identify effects for sub-groups that vary in terms of the alternative options available to them. 
\citet{hull2018isolateing} offers another approach to identifying principal effects based on an assumption of conditional homogeneity; \citet{jiang2016principal} derive a similar result for identifying principal effects based on experimental replications.  
The approach we use below is more agnostic than any of these, and focuses instead on partial identification. 
The idea is to see how far we can get in making sense of effect heterogeneity without having to impose strong and difficult-to-defend assumptions on the data generating process.  
In this way, the current paper is closest to \citet{miratrix2018bounding}, who study the Early College High School Initiative in the US. They also work with the same principal strata setup and derive bounds on the partially identified principal effects.  
Where we go beyond \citet{miratrix2018bounding} is to use a moment-based approach to formally justify both asymptotic-analytical and bootstrap-based inference for the bounds.  
Finally, our goal of unpacking heterogenous effects is similar to that of \citet{angrist2013explaining} in their analysis of heterogeneous effects for charter schools in the US, except that we take a principal strata approach rather than performing conditional-on-observables (e.g., Oaxaca-Blinder) decompositions.

The next section introduces our applied setting, which involves randomized controlled trial (RCT) replications, in 2008 and 2015, studying a community-based education (CBE) schooling intervention in rural Afghanistan.  
The interventions sought to increase primary education for girls in remote communities by establishing CBE schools inside the girls' home villages.
The two replications were six years apart, and during that time patterns in the rate of attendance in formal Afghan government schools changed substantially, even after we control for changes in the proximity of government schools.
The effects of introducing CBE on test scores were much stronger in the first RCT than in the second, possibly because of these changes in households' use of the government schools as an already-existing alternative.
To examine this possibility, we specify the relevant principal effects.  
We then develop nonparametric bounds on effects in two crucial principal strata: those who only attend formal school when a CBE school is introduced and those who switch from a government school to the CBE school.
We also develop methods for honest statistical inference with these bounds.  
Applying these methods to the Afghanistan RCTs, we find that the share of girls for whom CBE represented a novel way to gain access to school was much smaller in 2015 as compared to 2008. 
Nonetheless, the efficacy of CBE for girls was higher in 2015. 
Thus, even if CBE improved access for a diminishing share of girls (given that alternatives were increasingly available), those girls stood to draw especially strong benefit from it. 
At the margin for girls gaining new access to school, the value of CBE remained very high even if the overall number of immediate beneficiaries had declined.

\section{Applied Setting}

We are motivated by a pair of randomized controlled trials (RCTs) studying the impact of community-based education in rural Afghanistan.  
Community-based education programs organize primary school classes in available community spaces taught by trained community members. 
It is a strategy to extend access to a formal primary school curriculum in remote communities that otherwise are far from any formal schools.
The strategy is commonly used in settings where state capacity is weak, perhaps due to war \citep{burde2004weak, kirk2007home}.   

The first of the two RCTs that we study took place in Afghanistan's central highlands, in Ghor province, between 2007 and 2008. The results were reported in \citet{burde2013bringing}.  
The study documented large beneficial effects on math and literacy, particularly for Afghan girls.  
A meta-analysis of education interventions by the Abdul Latif Jameel Poverty Action Lab (JPAL) found this intervention to be among the most cost-effective ever evaluated with an RCT \citep{dhaliwal2013comparative}.
   
This success motivated a second RCT---a replication study that was fielded in the same geographic region  (although not exactly the same villages) between 2014 and 2015. Those results were reported in \citet{burde2016report}.  The replication also found strong beneficial effects, although the magnitudes were smaller than the previous experiment. 
This section characterizes those differences.

Of primary interest in these experiments was the effect of community-based schools on math and verbal ability of girls.
Girls are especially affected by the long distances between their home villages and already-existing government schools. 
In the longer working-paper version of their study, \citet[Fig. 6]{burde2012effect} showed that attendance rates were strongly correlated with school distance: attendance rates were about 80\% when a school was inside the village but then declined to below 50\% if the school was at least 2km away and then below 25\% when at least 3km away.
Needless to say, where attendance was low, math and verbal test scores were low as well.
Further analyses of surveys and qualitative evidence showed that this relationship between attendance and distance was likely due to parents being uncomfortable with having their daughters travel long distances to school \citep{burde2016report}. 
As a result, girls' math and verbal ability were strongly affected by the absence of a nearby government school. 

Using data from the two studies, we constructed a standardized math-verbal test score so as to allow for a comparison of effects on girls across the two studies.\footnote{We find strong positive effects on both math and verbal test scores, and these scores are positively correlated although not completely redundant, and so we combine them here to increase statistical power.} 
The test scores use either exactly the same questions or highly comparable questions, calibrated in terms of difficulty. 
We construct an index by, first, simply summing the number correct answers across both the math and verbal assessments and then standardizing. 
To ensure comparability, we standardized all scores with respect to the pooled mean and standard deviation from the 2008 RCT data.
Table \ref{itt-table} shows intention-to-treat effects of community-based education on the standardized test score. 
The first column of estimates are for the 2008 RCT, with a highly significant positive estimated effect of 0.73sd. The second column of estimates are based on reweighting the sample to account for differences between the 2008 RCT and the RCT ending in 2015 in sample households' proximity to the nearest government school.  
We discuss this more below, but for the moment we note that the reweighted sample estimate does not differ much from the unweighted sample estimate.
The third column of estimates is for the 2015 RCT, which we see is less than half (0.35sd) the size of the estimate from the 2008 RCT. 
The control means differ as well, although not by as much as the effects, particularly when comparing the reweighted 2008 sample to the 2015 sample.

\begin{table}[!t]
\centering
\begin{tabular}{lrrr}
   \hline
 & t=1 (2008) unweighted & t=1 (2008) weighted & t=2 (2015) \\ 
   \hline
ITT Effect & 0.73 & 0.78 & 0.35 \\ 
    & (0.11) & (0.14) & (0.13) \\ 
  Control mean & -0.36 & -0.23 & -0.17 \\ 
    & (0.09) & (0.14) & (0.12) \\ 
   \hline
N & 689 & 689 & 1218 \\ 
   \hline
\end{tabular}
\caption{Least squares regression estimates of the intention-to-treat (ITT) effects and control group means. Outcome is combined math-verbal test score, standardized. Standard errors accounting for village-level clustering in parentheses.} 
\label{itt-table}
\end{table}

\begin{figure}[htbp]
\begin{center}
\centerline{\includegraphics[width=.65\textwidth]{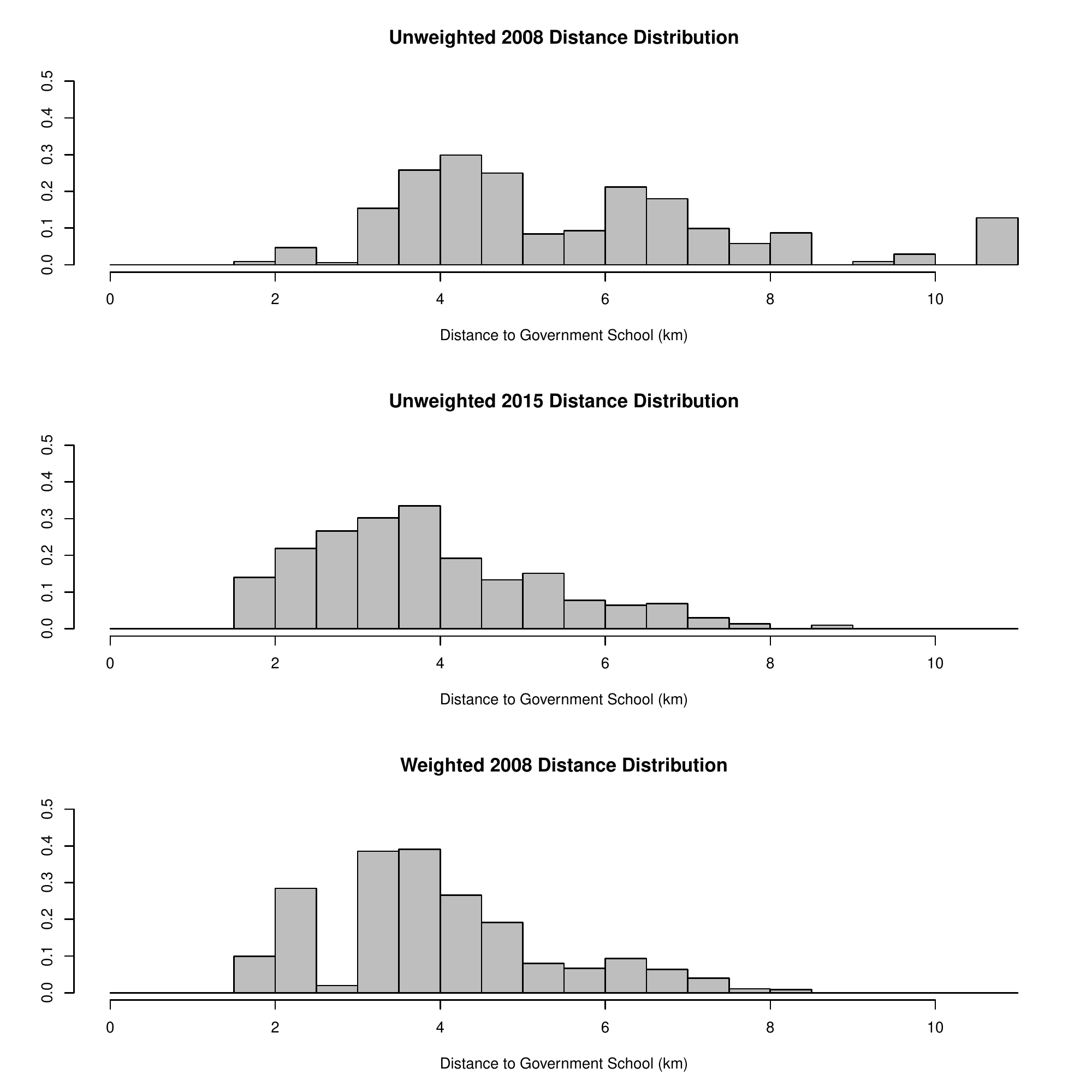} }
\caption{Distribution of households' distance to schools for the 2008 RCT, 2015 RCT, and then the 2008 RCT reweighted by the ratio of the 2015 and 2008 densities.}
\label{fig:distance-hist}
\end{center}
\end{figure}

What explains the difference?  
One might wonder if the population of children differs substantially across the two RCTs.  
In the analysis presented here, we are working with samples drawn from the same districts in Ghor province, Afghanistan, which is where the first RCT was conducted.  
But the two samples were taken in 2008 versus 2015.
In the interim, the Afghan government had made progress in extending the reach of the network of formal government schools.  
Thus, one important factor that is visibly different across the two replications is the distribution of distances between households and the nearest government schools. 
The distributions of households' distances to the nearest government schools for the two samples is displayed in Figure~\ref{fig:distance-hist}. 
The top panel shows a histogram of the distance of a child's home to the nearest government school in the 2008 sample. 
The middle panel shows the same histogram for the 2015 sample.  
Recall that \citet{burde2012effect} had found that attendance rates dropped off below 50\% when schools were at least 2km away, and below 25\% when schools were at least 3km away.  
The bulk of schools are at such a distance for all distributions shown in Figure~\ref{fig:distance-hist}, although we do see a non-negligible increase in the number of localities that are between 2-3km of a government school.
To account for this difference, the third panel shows the 2008 sample reweighted by the ratio of 2015 densities to 2008 densities \citep{sugiyama2012density}.
Such density ratio weighting balances the distribution of distances to the nearest government schools.
We used these balancing weights for the second column of estimates in Table \ref{itt-table}.
Again, we find that the ITT estimates differ substantially even after applying these balancing weights.

Perhaps there are other important differences at the household level? Table \ref{xsum-table} presents balance statistics across the two RCTs.  The middle column applies the density ratio weights to balance the distributions of households' distances to nearest government schools schools.  
Demographically, the children and their households do not differ dramatically.
The mean age of the children, household head's age, and household head's years of schooling are similar, particularly when considered relative to the standard deviations of these variables.
Somewhat imbalanced are the occupations of household heads, amount of land owned, and the ethnic distribution, measured in terms of the share of the sample that is Tajik (versus Hazara or Aimaq, which are the other ethnic groups that appear in the samples).
Also,  while in 2008, all programming was implemented by the same NGO in the sample, in 2015, 13\% of children had programs that were implemented by another NGO, although the intervention protocols were very similar. 

To evaluate whether these covariate differences matter, even if the differences are sometimes slight, Tables~\ref{itt-interactions-2008}~and~\ref{itt-interactions-2015} characterize effect heterogeneity with respect to covariates.  We see very little indication of effect heterogeneity with the exception of Tajik versus non-Tajik households in the 2008 sample.  In Column 6 of Table~\ref{itt-interactions-2008}, we see that when Tajik households are excluded, the estimated ITT is about 0.20sd smaller.  In Column 6 of Table~\ref{itt-interactions-2015} we see that the estimated ITT does not change when we limit attention to non-Tajik households in 2015, however, and when comparing across 2008 and 2015, a large difference still remains.

\begin{table}[!t]
\centering
\begin{tabular}{llrrr}
   \hline
Covariate &  & t=1 (2008) & t=1 (2008, weighted) & t=2 (2015) \\ 
   \hline
Child's age (in yrs.) & mean & 8.58 & 8.71 & 8.26 \\ 
   & (sd) & (1.56) & (1.58) & (1.51) \\ 
   &  &  &  &  \\ 
  Household head age (in yrs.) & mean & 40.61 & 39.99 & 45.03 \\ 
   & (sd) & (10.57) & (10.43) & (15.67) \\ 
   &  &  &  &  \\ 
  Household head yrs. of school & mean & 3.35 & 3.75 & 2.93 \\ 
   & (sd) & (3.39) & (3.86) & (2.97) \\ 
   &  &  &  &  \\ 
  Household head farmer (0/1) & mean & 0.77 & 0.74 & 0.56 \\ 
   & (sd) & (0.42) & (0.44) & (0.50) \\ 
   &  &  &  &  \\ 
  Land owned (jeribs) & mean & 1.13 & 1.07 & 0.69 \\ 
   & (sd) & (1.50) & (1.44) & (0.46) \\ 
   &  &  &  &  \\ 
  Tajik ethnicity (0/1) & mean & 0.26 & 0.23 & 0.11 \\ 
   & (sd) & (0.44) & (0.42) & (0.32) \\ 
   &  &  &  &  \\ 
  NGO from 2008 (0/1) & mean & 1.00 & 1.00 & 0.87 \\ 
   & (sd) & (0.00) & (0.00) & (0.34) \\ 
   &  &  &  &  \\ 
   \hline
N &  & 689 & 689 & 1218 \\ 
   \hline
\end{tabular}
\caption{Covariates means and standard deviations (in parentheses) for the t=1 (in 2008, both unweighted and weighted) and t=2 (in 2015) study populations.} 
\label{xsum-table}
\end{table}

\begin{table}
\begin{center}
\begin{scriptsize}
\begin{tabular}{l c c c c c c}
\hline
 & 1 & 2 & 3 & 4 & 5 & 6* \\
\hline
Control mean          & $-0.21$  & $-0.23$  & $-0.23$  & $-0.24$  & $-0.23$  & $-0.16$  \\
                      & $(0.12)$ & $(0.14)$ & $(0.14)$ & $(0.14)$ & $(0.13)$ & $(0.16)$ \\
Treatment             & $0.71$   & $0.78$   & $0.79$   & $0.78$   & $0.77$   & $0.58$   \\
                      & $(0.12)$ & $(0.14)$ & $(0.14)$ & $(0.14)$ & $(0.14)$ & $(0.16)$ \\
Child's age           & $0.29$   &          &          &          &          &          \\
                      & $(0.08)$ &          &          &          &          &          \\
Tr. X Child's age     & $0.09$   &          &          &          &          &          \\
                      & $(0.08)$ &          &          &          &          &          \\
HH. head age          &          & $-0.00$  &          &          &          &          \\
                      &          & $(0.00)$ &          &          &          &          \\
Tr. X HH. head age    &          & $0.01$   &          &          &          &          \\
                      &          & $(0.01)$ &          &          &          &          \\
HH. head school       &          &          & $0.03$   &          &          &          \\
                      &          &          & $(0.02)$ &          &          &          \\
Tr. X HH. head school &          &          & $0.02$   &          &          &          \\
                      &          &          & $(0.05)$ &          &          &          \\
HH. head farm.        &          &          &          & $-0.28$  &          &          \\
                      &          &          &          & $(0.31)$ &          &          \\
Tr. X HH. head farm.  &          &          &          & $0.29$   &          &          \\
                      &          &          &          & $(0.31)$ &          &          \\
Land                  &          &          &          &          & $-0.07$  &          \\
                      &          &          &          &          & $(0.05)$ &          \\
Tr. X Land            &          &          &          &          & $-0.06$  &          \\
                      &          &          &          &          & $(0.13)$ &          \\
\hline
N             & $688$    & $689$    & $688$    & $689$    & $689$    & $507$    \\
\hline
\end{tabular}
\end{scriptsize}
\caption{Least squares regression estimates of 2008 sample intention-to-treat and covariate-interaction effects. Outcome is combined math-verbal test score, standardized. Standard errors account for village-level clustering.
 *Model 6 omits Tajik households. }
\label{itt-interactions-2008}
\end{center}
\end{table}

\begin{table}
\begin{center}
\begin{scriptsize}
\begin{tabular}{l c c c c c c c}
\hline
 & 1 & 2 & 3 & 4 & 5 & 6* & 7** \\
\hline
Control mean          & $-0.19$  & $-0.18$  & $-0.20$  & $-0.18$  & $-0.18$  & $-0.21$  & $-0.19$  \\
                      & $(0.10)$ & $(0.12)$ & $(0.12)$ & $(0.12)$ & $(0.12)$ & $(0.13)$ & $(0.13)$ \\
Treatment             & $0.38$   & $0.36$   & $0.38$   & $0.35$   & $0.35$   & $0.38$   & $0.36$   \\
                      & $(0.11)$ & $(0.13)$ & $(0.13)$ & $(0.13)$ & $(0.13)$ & $(0.14)$ & $(0.15)$ \\
Child's age           & $0.26$   &          &          &          &          &          &          \\
                      & $(0.04)$ &          &          &          &          &          &          \\
Tr. X Child's age     & $0.00$   &          &          &          &          &          &          \\
                      & $(0.05)$ &          &          &          &          &          &          \\
HH. head age          &          & $0.00$   &          &          &          &          &          \\
                      &          & $(0.00)$ &          &          &          &          &          \\
Tr. X HH. head age    &          & $0.00$   &          &          &          &          &          \\
                      &          & $(0.00)$ &          &          &          &          &          \\
HH. head school       &          &          & $0.02$   &          &          &          &          \\
                      &          &          & $(0.01)$ &          &          &          &          \\
Tr. X HH. head school &          &          & $-0.00$  &          &          &          &          \\
                      &          &          & $(0.02)$ &          &          &          &          \\
HH. head farm.        &          &          &          & $-0.19$  &          &          &          \\
                      &          &          &          & $(0.11)$ &          &          &          \\
Tr. X HH. head farm.  &          &          &          & $0.07$   &          &          &          \\
                      &          &          &          & $(0.13)$ &          &          &          \\
Land                  &          &          &          &          & $-0.10$  &          &          \\
                      &          &          &          &          & $(0.15)$ &          &          \\
Tr. X Land            &          &          &          &          & $-0.01$  &          &          \\
                      &          &          &          &          & $(0.18)$ &          &          \\
\hline
N             & $1218$   & $1218$   & $1174$   & $1218$   & $1218$   & $1099$   & $1060$   \\
\hline
\end{tabular}
\end{scriptsize}
\caption{Least squares regression estimates of 2015 sample intention-to-treat and covariate-interaction effects. Outcome is combined math-verbal test score, standardized. Standard errors account for village-level clustering.
 *Model 6 omits Tajik households.
 **Model 7 restrits to communities with same NGO as 2008}
\label{itt-interactions-2015}
\end{center}
\end{table}

Much stronger differences are apparent when we look at school attendance patterns displayed in Table~\ref{attendance}.  Note that the estimates for 2008 incorporate the weights to balance the distributions in households' distances to nearest government schools across the 2008 and 2015 samples.  We see strikingly different levels of non-attendance in any formal school (``No school'') in the control groups across the two time periods (66\% in 2008 versus 36\% in 2015).  In part, this is attributable to higher rates of attendance at government schools (34\% in 2008 versus 45\% in 2015), perhaps attributable to improvements in security and other factors that make it easier to travel to government schools.  It is also attributable to apparent attendance of CBE classes for some control group households in 2008.  Given that we are controlling for households' distance from schools (as per Figure~\ref{fig:distance-hist}), it is clear that some households must be choosing in 2015 to send girls to attend school at distances that households were not willing to do in 2008.  
Improvements in the security condition, increased belief in the value of formal eduction, or other factors related to attendance decisions must be at work.

We also see strikingly different attendance patterns in the treatment groups.  In 2008, attendance among treated households is heavily concentrated in the CBE classes, whereas this is less so in 2015.  A full quarter of girls in the sample are attending government schools despite the existence of a CBE class in their village.  

\begin{table}[!t]
\centering
\begin{tabular}{|l|rr|rr|}
  \hline
  & \multicolumn{2}{c|}{2008 } &  \multicolumn{2}{c|}{2015} \\
 \hline
 & Control & Treated & Control & Treated \\ 
   \hline
No school & 0.66 & 0.23 & 0.36 & 0.20 \\ 
  Govt. school & 0.34 & 0.09 & 0.45 & 0.25 \\ 
  CBE & 0.00 & 0.67 & 0.18 & 0.54 \\ 
   \hline
\end{tabular}
\caption{Attendance rates for girls in control (no CBE) and treated (with CBE) communities in the t=1 (in 2008, estimated using balancing weights) and t=2 (in 2015) study populations.} 
\label{attendance}
\end{table}

Thus, we find strong behavioral differences across the two periods in terms of rates at which girls are accessing government school.  
This is clear when we look at attendance patterns in the control group.  
We also see differences across the two samples in the rate at which the presence of a CBE class induces girls to switch from the government school alternative to the CBE class.  
Such differences are presumably important factors in determining the ITT effects, and therefore these differences in attendance patterns may explain the differences in the ITT effects across the two samples. 
But these attendance patterns are post-treatment. 
Accounting for differences in the 2008 and 2015 attendance patterns is not something that can be done through a straightforward covariate control strategy.  
In the next section, we explain how we can use a principal stratum approach to account for such differences. 

\section{Formal Setting}

The analysis in the previous section revealed that attendance patterns varied dramatically across the two replications.
In this section, we develop an approach to link such attendance patterns to the ITT effects, which varied greatly across the replications. 
We decompose the ITT effects into effects for subgroups that attend CBE in the presence of the intervention but in the absence of the intervention, would either (1) not attend school at all or (2) attend a government school.  
These behavioral patterns define the principal strata for our analysis \citep{frangakis_rubin2002}, and effects that pertain to different principal strata are known as ``principal effects.'' 
We use general arguments about choice behavior and the importance of the curricular content of schooling to motivate assumptions that partially identify these principal effects. 
We derive corresponding estimators and inferential methods for the set-identified parameters \citep{manski1989anatomy, lee2009training, miratrix2018bounding}.  

We first define the formal setting.  We suppose that in period $t \in \{1,2\}$ we have a sample $\mathcal{U}_t$ of $N_t$ units indexed by $i \in \{1,...,N_t\}$. For a unit $i$ in period $t$, we have a treatment assignment $Z_{it} \in \{0,1\}$, and then potential school attendance outcomes under control ($Z_{it} = 0$) and treatment ($Z_{it} = 1$) defined as $(D_{it}(0),D_{it}(1)) \in \{ \{0,1,2\} \times  \{0,1,2\}\},$ with 0 meaning ``no school'',  meaning ``government school'', and 2 meaning ``CBE''.  Realized attendance outcomes are given by $D_{it} = Z_{it} D_{it}(1) + (1-Z_{it})D_{it}(0)$.

Table~\ref{attendtypes}~(1) enumerates the possible principal strata that could be defined with respect to the school attendance potential outcomes.   Given that there are three possible schooling attendance options in both treatment and control (no school, government school, or CBE), there are nine potential combinations for attendance under control and treatment.   We can remove some of these combinations from consideration under a revealed preference argument: if the introduction of CBE does not change the way ``no school'' is evaluated relative to ``government school,'' then a household should either stick with the same schooling choice for their child or switch their child to CBE \citep{kline2016bounding, heckman2018unordered}.  This argument can be formalized as a monotonicity assumption:
\begin{assn}[Monotonicity]\label{assn:monotonicity}
For all units in $\mathcal{U}_t$ and $t \in \{1,2\}$, if $D_{it}(1) \ne D_{it}(0)$, then $D_{it}(1) = 2$.
\end{assn}
\noindent Thus, either the CBE intervention has no effect on a given child, or it induces that child to attend CBE.  
Table~\ref{attendtypes}~(2) shows which principal strata remain relevant under monotonicity.  The cells labeled with ``X'' cannot be occupied if monotonicity holds.  The cells that can be occupied are given labels to characterize how treatment affects attendance patterns.  We use the variable $S_{it} = (D_{it}(0), D_{it}(1))$ to characterize the principal stratum for unit $i$.  The share of the population occupied by units for which $S_{it} = (a,b)$ is denoted as $\pi_t(a,b)$.  Table~\ref{attendtypes}~(3) shows how these terms are applied for the different principal strata.

Our ultimate interest is in the effects of the CBE intervention on learning, which we measure using test scores.  More generally, for each unit we have a potential outcome, $Y_{it}(d,z) \in \mathbb{R}$, under treatment assignment $Z_{it} = z$ and attendance outcome $D_{it} = d$, for all $d \in \{0,1,2\}$ and $z \in \{0,1\}$.  
Our analysis takes the curricular content of schooling seriously: we assume that the effect of having access to a school on learning depends on actually attending the school and that the presence or absence of the CBE intervention does not modify outcomes once effects on attendance are taken into account.
We formalize this in terms of the following exclusion restriction: 
\begin{assn}[Exclusion restriction]\label{assn:exclusion}
For all units in $\mathcal{U}_t$ and $t \in \{1,2\}$, $Y_{it}(d,z) = Y_{it}(d)$.
\end{assn}
As a result of the exclusion restrictions, observed outcomes are given by
$$
Y_{it} = \sum_{d \in \{0,1,2\}}Y_{it}(d)\mathbb{I}(D_{it}=d).
$$

Finally, motivated by the design of the RCTs, we assume conditional random assignment. Given observed covariates $X_{it} \in \mathcal{X}_t \subseteq \mathbb{R}^k$, we have:
\begin{assn}[Conditional random assignment]\label{assn:cia}
For all units in $\mathcal{U}_t$ and $t \in \{1,2\}$, 
$$
(D_{it}(z), Y_{it}(d,z)) \independent Z_{it} \mid X_{it} =x \text{ and } 0 < \text{Pr}[Z_{it} = 1 \mid X_{it} = x] < 1
$$ 
for all $d \in \{0,1,2\}$, $z \in \{0,1\}$, and $x \in \mathcal{X}_t$.
\end{assn}
\noindent For the RCTs that we study, conditional random assignment holds by construction, where the relevant covariates include the blocking variables that were used in the experimental design.  

\begin{table}
	\centering
\begin{scriptsize}
\begin{tabular}{c|c}
	\begin{tabular}{l c | c c c }
\multicolumn{5}{c}{\bf (1)}\\
\multicolumn{5}{c}{\bf All possible principal strata}\\
\hline
		& & 	\multicolumn{3}{c}{If assigned to treatment}\\
		& &  No school & Gov. School & CBE\\ 
		\hline	 \multirow{3}{*}{\rotatebox[origin=c]{90}{If assigned to control }} & \rotatebox[origin=c]{90}{\space\space\space No school\space\space\space }  &  &  &  \\
		& \rotatebox[origin=c]{90}{ Gov. School}  &  &  &  \\
		& \rotatebox[origin=c]{90}{\space\space\space CBE\space\space\space\space\space\space} &    &  &   \\
	\hline
	\end{tabular} 
& 
	\begin{tabular}{l c | c c c }
\multicolumn{5}{c}{\bf (2) }\\
\multicolumn{5}{c}{\bf Principal strata under monotonicity}\\
\hline
		& & 	\multicolumn{3}{c}{If assigned to treatment }\\
		& &  No school & Gov. School & CBE\\ 
		\hline	 \multirow{3}{*}{\rotatebox[origin=c]{90}{If assigned to control}} & \rotatebox[origin=c]{90}{\space\space\space No school\space\space\space } & Never taker & X & Complier \\
		& \rotatebox[origin=c]{90}{ Gov. School} & X & Gov. adherent & Substitutor \\
		& \rotatebox[origin=c]{90}{\space\space\space CBE\space\space\space\space\space\space} &   X & X & Always taker  \\
	\hline
	\end{tabular}
\vspace{1em}
\end{tabular}
\begin{tabular}{lcc}
\multicolumn{3}{c}{\bf (3) }\\
\multicolumn{3}{c}{\bf Characterizing principal strata under monotonocity }\\
\hline
&$ S_{it} =(D_{it}(0) , D_{it}(1))$ & Population share\\
\hline
 Never taker &(0, 0) & $\pi_t(0, 0)$\\
 Govt. adherent &(1, 1) & $\pi_t(1, 1)$\\
 Always taker & (2, 2)& $\pi_t(2, 2)$\\
 Complier & (0, 2)& $\pi_t(0, 2)$\\
 Substitutor & (1, 2)& $\pi_t(1, 2)$\\
 \hline
\end{tabular}
\end{scriptsize}
 \caption{The top left table (1) shows all possible principal strata given that there are three attendance possibilities in both control and treatment. The top right table (2) shows which principal strata could be potentially occupied under the monotonocity assumption. The bottom table (3) characterizes the principal strata in terms of the stratum variables and population shares.}\label{attendtypes}
\end{table}

In the analyses below, the population shares in the principal strata sometimes play a role.  Under monotonicity and conditional random assignment, the following hold:
\begin{align*}
\E[\E[\mathbb{I}(D_{it}=0) \mid Z_i = 1, X_{it}=x]] & = \pi_t(0,0) \\
\E[\E[\mathbb{I}(D_{it}=1) \mid Z_i = 1, X_{it}=x]] & = \pi_t(1,1) \\
\E[\E[\mathbb{I}(D_{it}=2) \mid Z_i = 0, X_{it}=x]] & = \pi_t(2,2) \\
\E[\E[\mathbb{I}(D_{it}=0) \mid Z_i = 0, X_{it}=x]] - \E[\E[\mathbb{I}(D_{it}=0) \mid Z_i = 1, X_{it}]] & = \pi_t(0,2) \\
\E[\E[\mathbb{I}(D_{it}=1) \mid Z_i = 0, X_{it}=x]] - \E[\E[\mathbb{I}(D_{it}=1) \mid Z_i = 1, X_{it}=x]] & = \pi_t(1,2).
\end{align*}
Thus, the principal strata shares are identified under monotonicity and conditional random assignment.  

Moreover, intention-to-treat (ITT) and the conventional instrumental-variables local average treatment effects (LATE) are identified in the usual ways \citep{angrist_etal96_late}.  For the ITT we have
\begin{align*}
\text{ITT} & = \E\left[Y_{it}(D_{it}(1), 1) - Y_{it}(D_{it}(0), 0) \right] \\
& = \E\left[\E\left[Y_{it}(D_{it}(1)) - Y_{it}(D_{it}(0)) \mid X_{it}=x\right]\right]\\
& = \E\left[\E\left[Y_{it} \mid Z_{it} = 1,X_{it}=x\right] - \E\left[Y_{it}\mid Z_{it}=0, X_{it}=x\right]\right]
\end{align*}
due to the exclusion restriction and conditional random assignment. For the LATE we have
\begin{align*}
\text{LATE} & = \E\left[Y_{it}(D_{it}(1), 1) - Y_{it}(D_{it}(0), 0) | D_{it}(1) \ne D_{it}(0) \right]  \\
& = \E\left[Y_{it}(2) - Y_{it}(D_{it}(0)) | D_{it}(1) \ne D_{it}(0) \right],
\end{align*}
due to the exclusion restriction and monotonicity.  Moreover by total probability and monotonicity,
\begin{align*}
\E\left[Y_{it} \mid Z_{it} = 1,X_{it}=x\right] & - \E\left[Y_{it}\mid Z_{it}=0, X_{it}=x\right] \\ & = \left\{\E [Y_{it}(2) - Y_{it}(D_{it}(0)) \mid D_{it}(1) \ne D_{it}(0), X_{it}=x]\right\}\\
& \hspace{10em} \times\text{Pr}[D_{it}(1) \ne D_{it}(0) \mid X_{it} = x],
\end{align*}
while,
$$
\text{Pr}[D_{it}(1) \ne D_{it}(0) \mid X_{it} = x] = \E [\mathbb{I}(D_{it}=2 \mid Z_{it}=1, X_{it}=x] - \E [\mathbb{I}(D_{it}=2 \mid Z_{it}=0, X_{it}=x].
$$
Thus under Assumptions \ref{assn:monotonicity}-\ref{assn:cia}, the LATE is identified with the usual ratio of the ITT and the ``first stage,'' where in this case the first stage refers to the effect of the intervention on CBE attendance rates.  The LATE  has the interpretation of the effect of inducing girls who would otherwise not attend school or attend government schools to attend CBE.

Finally, the following decomposition holds:
\begin{align*}
\text{LATE} & = \E\left[Y_{it}(2) - Y_{it}(0) | S_{it} = (0,2) \right]\frac{\pi_t(0,2)}{\pi_t(0,2) + \pi_t(1,2)} \\ 
& \hspace{2em} + \E\left[Y_{it}(2) - Y_{it}(1) | S_{it} = (1,2) \right]\frac{\pi_t(1,2)}{\pi_t(0,2) + \pi_t(1,2)}, 
\end{align*}
which means that the LATE is a linear combination of the principal effects for principal strata $S_{it} = (0,2)$ and $S_{it} = (1,2)$.  For any value of the LATE, which is point identified along with the principal strata shares under Assumptions \ref{assn:monotonicity}-\ref{assn:cia}, having a value for the $S_{it} = (0,2)$ principal effect implies that we can infer the $S_{it} = (1,2)$ effect, and vice versa.

\section{Bounds on Principal Effects}

In the absence of additional assumptions on the nature of the causal effects, the principal effects are not point identified.  It would be difficult, in this example, to defend additional assumptions beyond those that we have specified as Assumptions \ref{assn:monotonicity}-\ref{assn:cia}.  As such, we propose to use partial identification to construct sharp bounds on the principal effects.  

In the analysis that follows, we lighten the notation by omitting the conditioning on $X_{it}$.  The reader should bear in mind that all of the results below hold if we think of expectations operations as taking, first, the covariate-specific means and then, second, marginalizing over the target covariate distribution.

First, consider the average treatment effect for the compliers: 
$$
\E \left[Y_{it}(2) -  Y_{it}(0) | S_{it} = (0,2) \right]  = \E\left[Y_{it}(2) | S_{it} = (0,2) \right] - \E\left[ Y_{it}(0) | S_{it} = (0,2) \right]
$$
Under Assumptions \ref{assn:monotonicity}-\ref{assn:cia}, the second term on the right-hand side is identified as follows:
\begin{align*}
E\left[ Y_{it}(0) |S_{it}=(0,2)\right]  = & \frac{\pi_t(0,0)+\pi_t(0,2)}{\pi_t(0,2)}E\left[Y_{it}| Z_{it}=0, D_{it}=0\right] \\
& -\frac{\pi_t(0,0)}{\pi_t(0,2)}E\left[Y_{it}|Z_{it}=1, D_{it}=0\right].
\end{align*}
The first term on the right-hand side is not point identified under these assumptions, however.  Units for which we observe the $Y_{it}(2)$ among those assigned to control ($Z_{it} = 0$ and $D_{it} = 2$) are always takers, and units for which we observe $Y_{it}(2)$ in the group assigned to treatment ($Z_{it} = 1$ and $D_{it} = 2$) are a mix of always-takers, compliers, and substitutors.  Nonetheless, the mixture distribution for {\it complier and substitutor} treated outcomes is identified, given that we are able to partial off the always-takers values that we can observe among those assigned to control. More formally, we have,
\begin{align*}
F_{Y(2)}\left(y|S_{it} \in \{(0,2), (1,2)\}\right) & =  \frac{\pi_t(0, 2)+\pi_t(1, 2)+\pi_t(2, 2)}{\pi_t(0,2)+\pi_t(1,2)}F_{t,Y_{it}(2)}(y|S_{it} \in \{(0,2),(1,2),(2,2)\}) \\
& \hspace{1em} - \frac{\pi_t(2,2)}{\pi_t(0,2)+\pi_t(1,2)}F_{t,Y_{it}(2)}(y|S_{it} = (2,2)) \\
& =  \frac{\pi_t(0, 2)+\pi_t(1, 2)+\pi_t(2, 2)}{\pi_t(0,2)+\pi_t(1,2)}F_{t,Y_{it}(2)}(y|Z_{it}=1, D_{it}=2) \\
& \hspace{1em} - \frac{\pi_t(2,2)}{\pi_t(0,2)+\pi_t(1,2)}F_{t,Y_{it}(2)}(y|Z_{it}=0, D_{it}=2)
\end{align*}

\begin{figure}[!t]
\begin{center}
\centerline{\includegraphics[width=.65\textwidth]{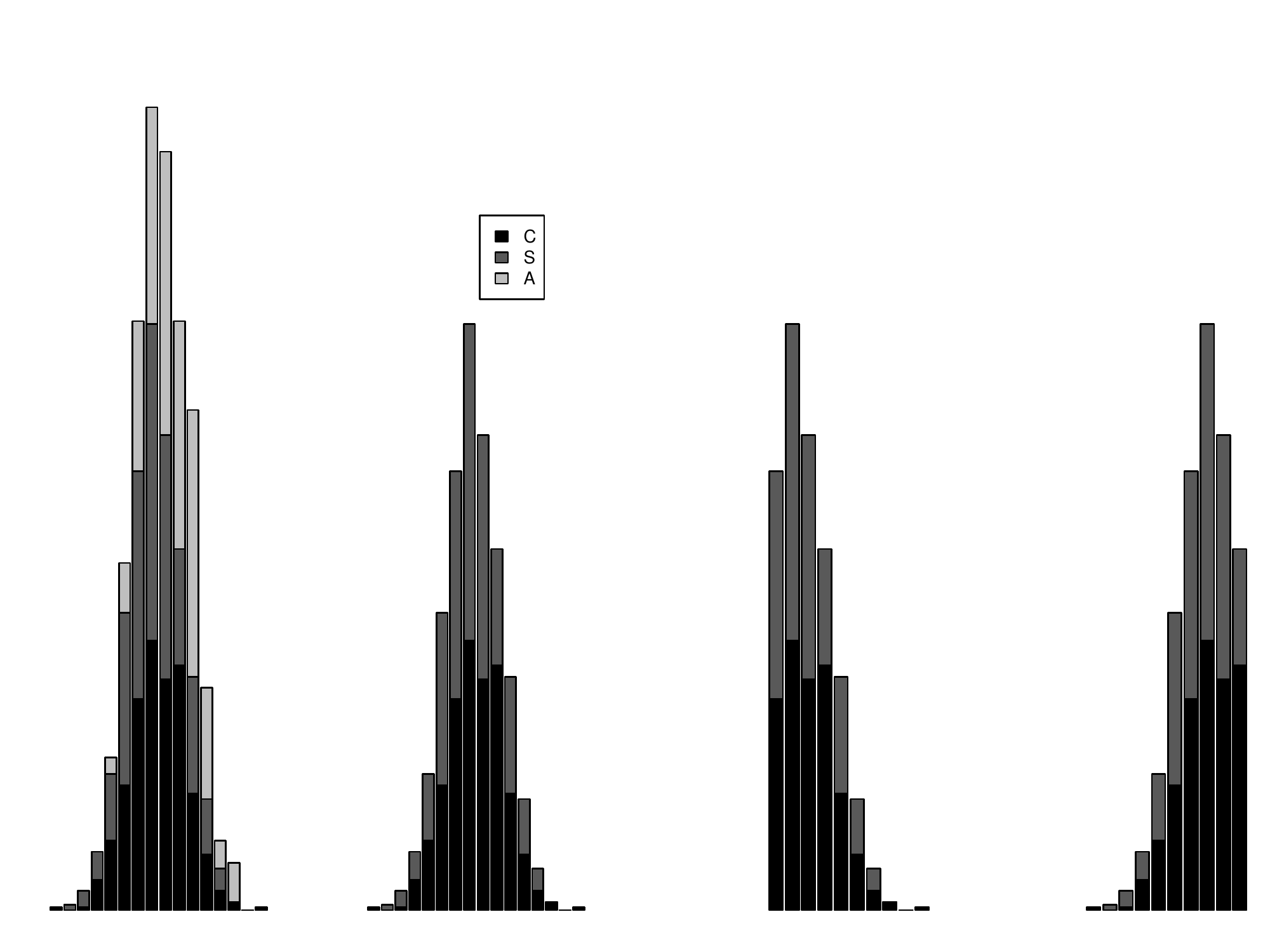} }
\caption{Suppose the left-most histogram is the marginal distribution of those assigned to treatment.  The outcome distribution of {\it always-takers} can be identified from the control group and subtracted from the treatment-group outcome distribution. Doing so leaves a mixture distribution of {\it complier} and {\it substitutor} outcomes, as shown in the second-to-left histogram. In the data, we cannot distinguish {\it compliers} from {\it substitutors} in the treatment group. But we can use the estimated share of substitutors to trim this mixture distribution from below (third-to-left) and above (right-most), allowing us to bound the mean of {\it complier} potential outcomes under treatment.}
\label{fig:trim}
\end{center}
\end{figure}

Recall from the discussion in the previous section that the share $\pi_t(1,2)$ of substitutors is identified. In a manner similar to \citet{lee2009training}'s bounds for intensive margin effects, we can use the estimated $\pi(1,2)$ share to ``trim'' the distribution $F_{Y_{it}(2)}\left(y|S_{it} \in \{(0,2), (1,2)\}\right)$ so as to bound features of the distribution for compliers, $F_{Y_{it}(2)}\left(y|S_{it} = (0,2) \right)$. This includes bounding the complier mean $\E\left[Y_{it}(2) | S_{it} = (0,2) \right]$ and, in turn, the complier average causal effect, $\E \left[Y_{it}(2) -  Y_{it}(0) | S_{it} = (0,2) \right]$.  Figure~\ref{fig:trim} provides a hypothetical illustration of the approach.

Finally, recall from the discussion in the previous section that the LATE is a linear combination of the complier and substitutor average causal effects.  As a result, a bound on the complier average causal effect in turn establishes a bound on the substitutor causal effect, $\E\left[Y_{it}(2) - Y_{it}(1) | S_{it} = (1,2) \right]$. 

The appendix provides details on implementing these ideas using either trimmed means or  a kernel density approach.  For inference, we work with standard methods for the ITT and LATE.
For the bounds, under regularity conditions similar to \citet{lee2009training}, the appendix establishes asymptotic normality and a method for approximating the asymptotic distribution. Asymptotic normality justifies use of the non-parametric bootstrap, and the bootstrap generally offers refinement over using the asymptotic approximation to construct confidence intervals \citep{horowitz-bootstrap-handbook}.

\section{Applied Results}

We now apply the principal strata methods to interpret the variation in effects across the two replications of the schooling RCT in Afghanistan.  
We recall from Table~\ref{itt-table} that the ITT effect on girls' test scores from the 2015 replication was less than half that of the ITT from 2008.
This difference in effect sizes was not clearly attributable to differences in locality or household characteristics.  
This difference in effect sizes remained even after accounting for changes in the proximity of government schools.
But from Table~\ref{attendance} we noted that control-group school attendance patterns differed markedly. 
We saw indication of change over time in government-school attendance behavior, even after controlling for distance to government schools.
We also saw a change in the extent to which households chose to switch their girls from government schools to CBE classes.  

Table~\ref{late-table} shows estimates of the LATE on girls' test scores, which captures the effect of attending CBE classes rather than {\it either} no schooling or government schooling (i.e., the LATE pools the compliers and substitutors).  
Assuming that households optimize in their school choices, the LATE identifies the effect of CBE relative to a household's next-best alternative, and therefore provides a summary measure of the welfare effects of the intervention \citep{kline_walters2016}.
The result here is remarkable: the LATEs are almost identical even though in 2008, most of the switching was from no school to CBE, whereas in 2015, a sizable fraction of those who attended CBE would have attended government school under control.  
Table~\ref{attendrates} captures these attendance pattern differences more precisely by showing the estimated distribution over principal strata in the weighted 2008 sample and 2015 sample.  
In 2008, those induced to take up CBE were about 63\% compliers and 37\% substitutors, whereas in 2015, those induced to take up CBE were about 44\% compliers and 56\% substitutors.

These patterns raises the possibility that the beneficial effects of CBE on test scores has {\it increased} for 2015 compliers as compared to 2008 compliers.
This could be because those who remained as compliers by 2015 (only 16\% of the population as compared to 42\% in 2008) actually {\it stood to gain more} than those who were newly accessing government school.  
If this were true, then it would imply that the {\it marginal} complier effect of CBE is  higher in 2015, even though the share of people who stand to benefit is smaller.
Such a phenomenon would be similar to the negative selection-on-gains patterns for early childhood education in the US \citep{kline_walters2016, cornelissen2018benefits}: in those studies, children who were unlikely to enroll in pre-school in the absence of the intervention appeared to gain the most when the intervention was introduced.  
Something similar could be at work in our example: those who did not take up expanded access to government school appear to benefit most from expanded access through CBE.
But the patterns also raise the possibility that CBE classes are {\it less} beneficial for substitutors in 2015 as compared to 2008.  
This is a simple consequence of the linear relationship between the LATE, complier effects, and substitutor effects.  

\begin{table}[!t]
\centering
\begin{tabular}{lrrr}
   \hline
 & t=1 (2008) unweighted & t=1 (2008) weighted & t=2 (2015) \\ 
   \hline
LATE & 1.06 & 1.06 & 1.04 \\ 
    & (0.14) & (0.22) & (0.53) \\ 
   \hline
First stage & 0.69 & 0.73 & 0.35 \\ 
    & (0.04) & (0.03) & (0.07) \\ 
   \hline
N & 1218 & N & 1218 \\ 
   \hline
\end{tabular}
\caption{Two-stage least squares regression estimates of the LATE on test scores, and least squares regression estimates of the first stage effect on CBE take-up. Standard errors accounting for village-level clustering in parentheses.} 
\label{late-table}
\end{table}

\begin{table}[!t]
\centering
\begin{tabular}{lrr}
   \hline
Principal stratum & t=1 & t=2 \\ 
   \hline
Never taker & 0.23 & 0.20 \\ 
  Govt. adherent & 0.09 & 0.25 \\ 
  Always taker & 0.00 & 0.18 \\ 
  Complier & 0.42 & 0.16 \\ 
  Substitutor & 0.25 & 0.20 \\ 
   \hline
\end{tabular}
\caption{Weighted estimation distribution of girls population across principal strata in the t=1 (in 2008) and t=2 (in 2015) study populations, assuming monotonicity.  Shares may not add up to 1 due to rounding.} 
\label{attendrates}
\end{table}

We dig into these possibilities with estimates for the principal effects for compliers and substitutors.  
By the arguments above, these effects are partially identified under (i) monotonicity in attendance behavior, (ii) an exclusion restriction such that the effect of the intervention on test scores is fully mediated by attendance effects, and (iii) the conditionally random assignment (i.e. Assumptions \ref{assn:monotonicity}-\ref{assn:cia}).  
We can use the bounds to evaluate the possibility of {\it negative} learning effects for those who switch from government school to CBE.  
We can also evaluate whether learning effects for compliers in 2015 may have been stronger than for compliers in 2008.  

\begin{figure}[htbp]
\begin{center}
\centerline{\includegraphics[width=.65\textwidth]{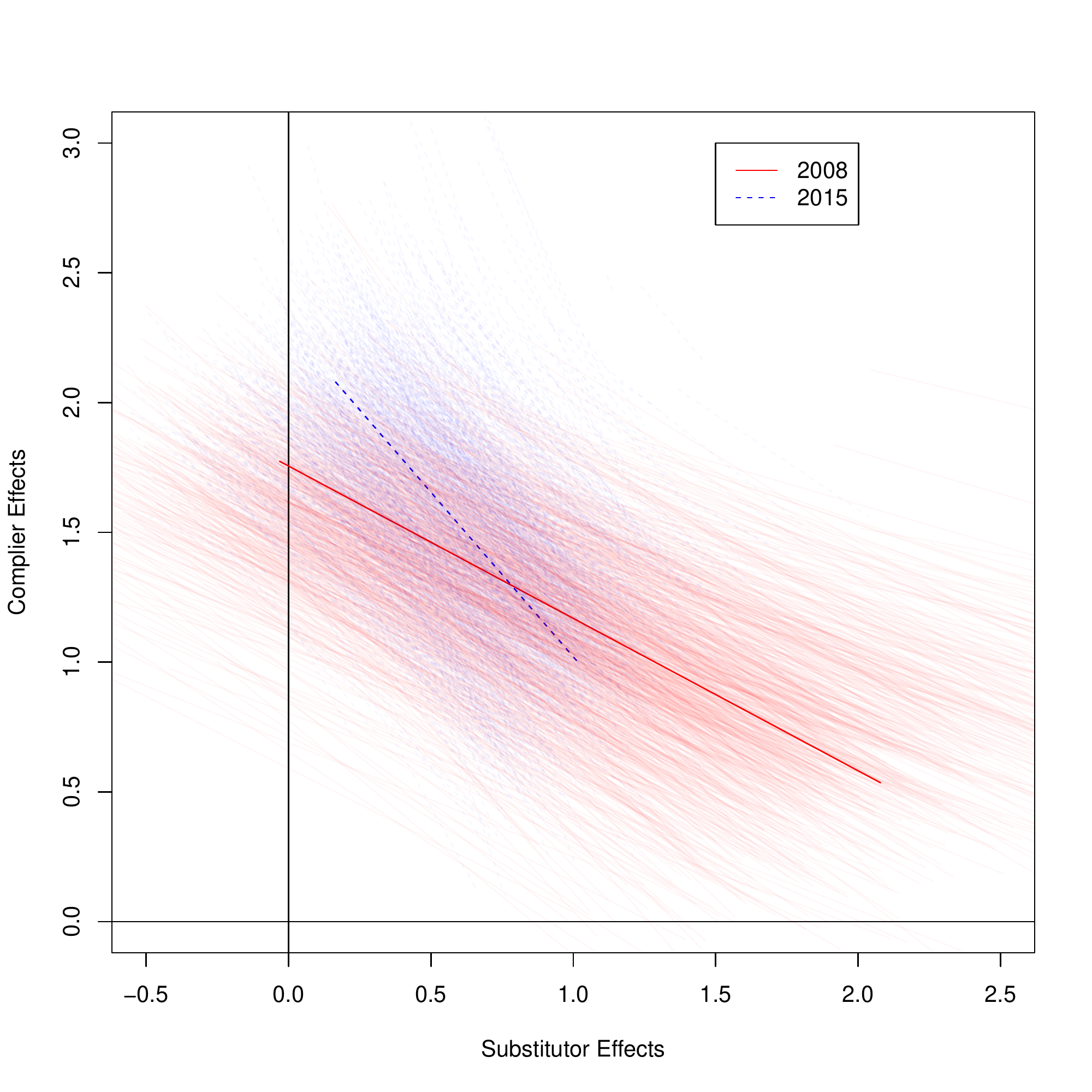} }
\caption{The red solid and blue dashed lines show the range of possible values for the effect on learning for substitutors and compliers in 2008 and 2015, respectively.  The linear relationship between the effects for substitutors and compliers and compliers is based on the linear decomposition of the LATE. The light streaks in the background show the bootstrap distributions for the bounds.}
\label{fig:bounds-plot}
\end{center}
\end{figure}

Our bounds estimates are displayed in Table~\ref{bounds_boot-table} and Figure~\ref{fig:bounds-plot}.
Recall that the LATE decomposes linearly into effects for substitutors and compliers. 
That linear relationship explains the way the bounds appear in Figure~\ref{fig:bounds-plot}.
Fixing the value of the LATE as per Table~\ref{late-table}, a given value for the substitutor effect implies a value for the complier effect. 
The lines on Figure~\ref{fig:bounds-plot} trace out the possible combinations that are consistent with the data.  
For inference, we also display non-parametric bootstrap confidence intervals in Table~\ref{bounds_boot-table}.
The faint streaks in Figure~\ref{fig:bounds-plot} are estimates on the bootstrap samples and allow us to visualize the uncertainty in our bounds estimates. 

First, consider the effects of CBE on compliers---girls between 6-11 years old that would have not attended any school in the control condition but attend CBE when it is available. 
In 2008, compliers made up the largest principal stratum at 42\% of girls, but in 2015, only 16\% of girls were in this category.
The 2008 RCT suggests strong positive effects for the compliers, between a 0.54 to 1.77 standard deviation average effect on test scores.  
The 2015 RCT suggests even stronger effects, with bounds between 0.99 and 2.08 standard deviations. 
The overlap between the 2008 and 2015 bounds means that we cannot rule out that effects are the same, but they also allow the possibility that effects for compliers were substantially stronger in 2015 than in 2008.
This could be due to improvements in the ways that the schools and teachers taught, although by our understanding the CBE protocols did not vary dramatically between 2008 and 2015.
In both cases, the classes were led by teachers recruited from the local community who were trained on the government curriculum, which also did not change dramatically.  
Given this similarity in the nature of the CBE intervention, another possibility is that the population of compliers in 2015, who were a small minority share of the overall population of girls, stood to gain more from CBE than the population of compliers in 2008, which included a much larger segment of the population of girls. 
Most likely any increase in efficacy for the compliers is due to a combination of these two factors, although by our understanding the effect of improved teaching is likely to be more limited.

Now, we could extend the analysis under an {\it assumption} of monotonicity in the relationship between selection into schooling and gains; doing so would allow us to characterize marginal treatment effects more precisely \citep{heckman1999local, kowalski2016doing}. 
Such a monotonicity assumption is not compelling, however. 
We observe possibly negative selection on gains from schooling {\it on average}. 
But this does not stand up to scrutiny as a structural assumption that we ought to presume for all households.
Presumably the usual positive selection-on-gains applies to some households, but the standard pattern is dominated by other forms of selection behavior across the 2008 and 2015 results.

Next, consider the effects on substitutors.
Presumably the reason that households would switch from having their daughters attend government school to attending CBE school is because it would allow them to attend school in their home village rather than having to travel to the government school outside their village.
In doing so, households may be trading possible learning benefits for the convenience of going to the CBE school in the village.
Such concerns are due to the fact that CBE classes are taught by community members who may not have the formal training that government schools require of their teachers.
Such a trade-off may nevertheless be welfare improving, if it is the case that learning loss is minor relative to the benefits of avoiding travel for school; see \citet{kline_walters2016} for a related discussion of such substitution effects. 
The 2008 estimates for effects on substitutors allow the possibility for negative effects. 
The point estimate for the bounds suggests this possibility is only slight, but the 95\% confidence interval admits the possibility of substantial negative effects. 
The upper bound, however, admits the possibility strong positive effects. 
The 2008 RCT results does not speak clearly about whether this choice also came with a learning benefit or cost.
The 2015 RCT results are clearer in this regard. 
The bounds suggest positive effects, although the confidence interval on the lower bound does allow for medium-sized negative effects.  
Thus, in 2015 at least, we can rule out substantially negative effects.  
The bounds also admit substantially positive effects.
Our interpretation is that in 2015, there was little reason to be concerned that households were trading off on their daughters' learning by having them switch from a government school to the more conveniently located CBE class.
This is all the more important in 2015, because substitutors constitute the largest share of CBE participants (as per Table~\ref{attendrates}).

\begin{table}[!t]
\centering
\begin{tabular}{lrrrr}
   \hline
Year & 2008 & 2008 & 2015 & 2015 \\ 
 & Lower bound, t=1 & Upper bound, t=1 & Lower bound, t=2 & Upper bound, t=2 \\ 
   \hline
  Compliers & 0.54 & 1.77 & 0.99 & 2.08 \\ 
  & (0.24,0.83) & (1.46,2.09) & (0.52,1.46) & (1.60,2.56) \\ 
 \hline
  Substitutors & -0.03 & 2.08 & 0.17 & 1.02 \\ 
  & (-0.65,0.59) & (1.30,2.86) & (-0.29,0.62) & (0.61,1.44) \\ 
   \hline
\end{tabular}
\caption{Bound estimates of the effect on compliers and substitutors. Bootstrap 95\% confidence interval in parentheses.} 
\label{bounds_boot-table}
\end{table}

We note that the lines characterizing the 2008 and 2015 bounds cross at around 0.75sd for the substitutor effect and 1.25sd for the complier effect. 
Were it the case that effects were homogenous in 2008 and 2015 for substitutors and compliers, this intersection offers a way to point identify these principal effects.
Such an approach is similar to that of \citet{hull2018isolateing} and \citet{jiang2016principal}. 
We resist such an interpretation, because the substitutor and complier subpopulations are clearly very different between 2008 and 2015.  
This is apparent by their large differences in the sizes of these principal strata (Table~\ref{attendrates}).

Our results helps us to understand why the intervention was substantially less effective, on average, in 2015 as compared to 2008.  
The main part of the story was that the 2015 share of compliers---those who would not attend school in the absence of the CBE intervention---was almost 1/3 the share in 2008. 
That difference was potentially offset by increases in efficacy for compliers.
In the absence of such an increase in efficacy we would expect the average effectiveness to be even smaller in 2015.

Our results also have important policy implications.
Our data strongly suggest a high value to marginal increases in access to schools via CBE.
That is, the comparison between 2008 and 2015 shows that as CBE was extended to cover people who were ever-less likely to attend any school in the absence of CBE, the efficacy of CBE increased.  
Even if CBE improves access to schools for a minority share of students, it seems those students stand especially to benefit from it, and so it may indeed be worth it to continue with the intervention. 

\section{Conclusion}

Interventions that offer new programs to individuals, such as new forms of schooling, typically vary in their effectiveness across contexts. 
To determine whether interventions should be extended in different contexts, it is crucial to understand why effectiveness varies. 
Variation in effectiveness may be because contexts vary in terms of the ways that individuals use already-existing alternatives to the intervention.
If so, contexts may vary in how the intervention changes use patterns, and this in turn would determine for whom the intervention may offer any benefits. 
Understanding such variation in use patterns is crucial, but it is also complicated by the fact that it requires understanding post-treatment differences that are not point identified by random assignment.
In this paper, we show that when a monotonicity and exclusion restriction also hold, we can partially identify effects for subgroups---principal strata---who differ in terms of how the intervention would affect their usage of the program relative to existing alternatives. 
We develop an approach to estimating bounds on the partially identified principal strata effects and doing inference on those bounds.

We apply these methods to analyzing 2008 and 2015 replications of a randomized controlled trial testing a community-based education (CBE) intervention in the same remote areas of Afghanistan.
Our analysis focuses on the benefits of CBE for girls in these remote villages.
While the average effects of CBE in 2015 were substantial, they were about half as large as in  2008.
The differences are not clearly attributable to observable characteristics of households or communities.
The main difference is in the ways that households use the existing alternatives.
Whether for reasons of safety or perceived value of schooling, households in 2015 were much more likely to send their daughters to government schools located outside their home villages. 
This meant that the share of girls who would attend no school in the absence of CBE was much smaller in 2015.
However, our evidence is consistent with CBE being much more efficacious for girls that would otherwise not attend school at all.
Therefore, even though the share of girls who newly gain access to school through CBE is much smaller in 2015 versus 2008, potentially much higher efficacy means that it may make sense to continue with the intervention.

The methods that we present here could be useful in other cases bear a structural resemblance to the case of CBE in Afghanistan---namely, situations where an intervention offers a new option alongside already-existing alternatives. 
There are limits, however. 
The case that we studied here was one in which there were two alternatives to the newly-introduced CBE program: one substitute form of schooling (government school) and one alternative that consisted of no schooling at all.
As the number of alternatives increases, the number of potential principal strata increases multiplicatively, and the restrictions required to even partially identify principal strata effects may become untenable.
Nonetheless, many situations are such that an intervention faces one dominant already-existing alternative, and for those cases, the methods here are generally applicable.

\clearpage

\section*{Appendix}

\begin{appendix}

\section{Deriving bounds on the principal effects}
The following proposition presents the expressions of the bounds and shows that the bounds are sharp for the conditional expectation of interest.
Let's denote the complier average causal effect, $\E \left[Y_{it}(2) -  Y_{it}(0) | S_{it} = (0,2) \right]$, as $\tau_t(0,2;(0,2))$, and the substitutor causal effect, $\E\left[Y_{it}(2) - Y_{it}(1) | S_{it} = (1,2) \right]$, as $\tau_t(1,2;(1,2))$, then
\begin{prop}\label{prop:bounds}
Suppose Assumptions \ref{assn:monotonicity}-\ref{assn:cia} hold and $Y_{it}(d)$ is continuous for all $d \in \mathcal{D}$. Then,
\begin{enumerate}[(i)]
\item $$
\beta_{t,L}(0,2) \le E\left[Y_{it}(2)|S_{it}=(0,2)\right] \le \beta_{t,U}(0,2),
$$ 
where
\begin{align*}
\beta_{t,L}(0,2) = & \frac{\pi_t(0,2)+\pi_t(1,2)}{\pi_t(0,2)}  \int_{-b_y}^{c_{t,1}} y d \tilde{F}_{t,Y_{it}(2)}(y)
\end{align*}
and
\begin{align*}
\beta_{t,U}(0,2) = & \frac{\pi_t(0,2)+\pi_t(1,2)}{\pi_t(0,2)}  \int^{b_y}_{c_{t,2}} y d \tilde{F}_{t,Y_{it}(2)}(y),
\end{align*}
with 
\begin{align*}
\tilde{F}_{t,Y_{it}(2)}(y) 
&=  \frac{\pi_t(0,2)+\pi_t(1,2) +\pi_t(2,2)}{\pi_t(0,2)+\pi_t(1,2)}F_{t,Y_{it}(2)}(y|Z_{it}=1, D_{it}=2) \\
& \hspace{2em} - \frac{\pi_t(2,2)}{\pi_t(0,2)+\pi_t(1,2)}F_{t,Y_{it}(2)}(y|Z_{it}=0, D_{it}=2),\\
c_{t,1} & \equiv \tilde{F}_{t,Y_{it}(2)}^{-1}\left(\frac{\pi_t(0,2)}{\pi_t(0,2)+\pi_t(1,2)} \right), \text{ and}\\
c_{t,2} & \equiv \tilde{F}_{t,Y_{it}(2)}^{-1}\left(\frac{\pi_t(1,2)}{\pi_t(0,2)+\pi_t(1,2)} \right).
\end{align*}
where $F_{t,V(v)}$ is the distribution function in context $t$ for a random variable $V$ and $F_{t,V}^{-1}(r)$ gives the quantile at point $r$ in the distribution function.
\item The bounds $[\beta_{t,L}(0,2), \beta_{t,U}(0,2)]$ are sharp for $E\left[Y_{it}(2)|S=(0,2)\right]$ with respect to the mixture distribution, $F_{t,Y_{it}(2)}\left(y|S_{it} \in \{(0,2), (1,2)\}\right)$.
\item 
$$
\beta_{t,L}(0,2) - \mu^0_t(0,2) \le \tau_t(0,2;(0,2)) \le \beta_{t,U}(0,2) - \mu^0_t(0,2).
$$
\end{enumerate}
\end{prop}

\noindent{\it Proof.} We cannot disentangle the distribution of $Y_{it}(2)$ outcomes for $S_{it} = (0,2)$ and $S_{it} = (1,2)$ types. Rather, these distributions are combined in the mixture distribution,
\begin{align}
F_{t,Y_{it}(2)}\left(y|S_{it} \in \{(0,2), (1,2)\}\right) = & \frac{\pi_t(0,2)}{\pi_t(0,2)+\pi_t(1,2)}F_{t,Y_{it}(2)}\left(y|S_{it} = (0,2)\right) \nonumber\\
&  + \frac{\pi_t(1,2)}{\pi_t(0,2)+\pi_t(1,2)}F_{t,Y_{it}(2)}\left(y|S_{it}=(1,2) \right).\label{eq:mix}
\end{align}
Under the assumptions, the mixture distribution is identified  by the data as,
\begin{align*}
F_{t,Y_{it}(2)}\left(y|S_{it} \in \{(0,2), (1,2)\}\right) & =  \frac{\pi_t(0, 2)+\pi_t(1, 2)+\pi_t(2, 2)}{\pi_t(0,2)+\pi_t(1,2)}F_{t,Y_{it}(2)}(y|S_{it} \in \{(0,2),(1,2),(2,2)\}) \\
& \hspace{1em} - \frac{\pi_t(2,2)}{\pi_t(0,2)+\pi_t(1,2)}F_{t,Y_{it}(2)}(y|S_{it} = (2,2)) \\
& =  \frac{\pi_t(\cdot, 2)}{\pi_t(0,2)+\pi_t(1,2)}F_{t,Y_{it}(2)}(y|Z_{it}=1, D_{it}=2) \\
& \hspace{1em} - \frac{\pi_t(2,2)}{\pi_t(0,2)+\pi_t(1,2)}F_{t,Y_{it}(2)}(y|Z_{it}=0, D_{it}=2)\\
& = \tilde F_{t,Y_{it}(2)}(y).
\end{align*}
Now consider $\beta_{t,L}(0,2)$ as defined above.  As is clear from expression \eqref{eq:mix}, no subpopulation with proportion $\frac{\pi_t(0,2)}{\pi_t(0,2)+\pi_t(1,2)}$ from the population distributed as $F_{t,Y_{it}(2)}\left(y|S_{it} \in \{(0,2), (1,2)\}\right)$ could have a mean smaller than $\beta_{t,L}(0,2)$, given that $\beta_{t,L}(0,2)$ concentrates all values into the bottom left of the distribution. This establishes $\beta_{t,L}(0,2) \le E\left[Y_{it}(2)|S=(0,2)\right]$, and a similar argument establishes $ \beta_{t,U}(0,2) \ge E\left[Y_{it}(2)|S=(0,2)\right]$, proving (i). For (ii), sharpness of the bounds follows from \citet[Corollary 4.1]{horowitz1995identification} as applied to expression \eqref{eq:mix} and the fact that the cut point, $c_{t}$, is uniquely determined by continuity of $Y_{it}(2)$ and the uniqueness of $\frac{\pi_t(0,2)}{\pi_t(0,2)+\pi_t(1,2)}$ . Then (iii) follows from the expressions for $\mu_t^0(0,2)$ and $\tau_t(0,2;(0,2))$.\hspace{1em}$\square$

We have similar results for $\tau_t(1,2;(1,2))$, which are omitted to save space.

\section{Estimation}

In this section, we present estimators for each parameter of interest mentioned in the main text. First, let's define $p(d, z) = P(D_{it} = d, Z_{it} = z)$. Then, we estimate the share of each principal strata as follows:
\begin{align*}
& \hat \pi_t(1,2) = \hat p_t(1,0) - \hat p_t(1,1) \\
= & \frac{\sum_{i=1}^{N_t} \frac{1-Z_{it}}{1-e(X_{it})}\If(D_{it} = 1)}{\sum_{i=1}^{N_t} \frac{1-Z_{it}}{1-e(X_{it})}} - \frac{\sum_{i=1}^{N_t} \frac{Z_{it}}{e(X_{it})}\If(D_{it} = 1)}{\sum_{i=1}^{N_t} \frac{Z_{it}}{e(X_{it})}} \\
& \hat \pi_t(2,2) = \hat p_t(2,0) = \frac{\sum_{i=1}^{N_t} \frac{1-Z_{it}}{1-e(X_{it})}\If(D_{it} = 2)}{\sum_{i=1}^{N_t} \frac{1-Z_{it}}{1-e(X_{it})}} \\
& \hat \pi_t(0,0) = \hat p_t(0,1) = \frac{\sum_{i=1}^{N_t} \frac{Z_{it}}{e(X_{it})}\If(D_{it} = 0)}{\sum_{i=1}^{N_t} \frac{Z_{it}}{e(X_{it})}} \\
& \hat \pi_t(0,2) = \hat p_t(0,0) - \hat p_t(0,1) \\
= & \frac{\sum_{i=1}^{N_t} \frac{1-Z_{it}}{1-e(X_{it})}\If(D_{it} = 0)}{\sum_{i=1}^{N_t} \frac{1-Z_{it}}{1-e(X_{it})}} - \frac{\sum_{i=1}^{N_t} \frac{Z_{it}}{e(X_{it})}\If(D_{it} = 0)}{\sum_{i=1}^{N_t} \frac{Z_{it}}{e(X_{it})}}
\end{align*}

Next, for the intention to treat effect (ITT) and local average treatment effect (LATE), we have
\begin{align*}
\widehat{\text{ITT}}_t \equiv \frac{\sum_{i=1}^{N_t} Y_{it} \frac{Z_{it}}{e(X_{it})}}{\sum_{i=1}^{N_t}\frac{Z_{it}}{e(X_{it})}} - \frac{\sum_{i=1}^{N_t} Y_{it} \frac{1-Z_{it}}{1-e(X_{it})}}{\sum_{i=1}^{N_t}\frac{1-Z_{it}}{1-e(X_{it})}},
\end{align*}

\begin{align*}
\widehat{\text{LATE}}_t \equiv \frac{\widehat{\text{ITT}}_t }{\hat \pi_t(0,2)+\hat \pi_t(1,2)}.
\end{align*}

Then,  for the lower bound of $E\left[Y_{it}(2)|S_{it}=(0,2)\right]$, $\beta_{t,L}(0,2)$, defined in the previous section, we estimate it as follows:
\begin{align*}
\hat \beta_{t,L}(0,2) \equiv \frac{\hat \pi_t(0,2)+\hat \pi_t(1,2)}{\hat  \pi_t(0,2)} \left[ \hat \pi_{t1}\hat \mu_t(-b_y,\hat c_{t,1}; 1, 2)  -  \hat \pi_{t2}\hat \mu_t(-b_y, \hat c_{t,1}; 0, 2) \right]
\end{align*}
where $\hat \pi_{t1} = \frac{\hat \pi_t(0,2)+\hat \pi_t(1,2)+\hat \pi_t(2,2)}{\hat \pi_t(0,2)+\hat \pi_t(1,2)}$, $\hat \pi_{t2} = \frac{\hat \pi_t(2,2)}{\hat \pi_t(0,2)+\hat \pi_t(1,2)}$, and
$$
\hat \mu_t(u,v; z, d) \equiv \frac{\sum_{i=1}^{N_t} \left[z\frac{Z_{it}}{e(X_{it})} + (1-z)\frac{1-Z_{it}}{1-e(X_{it})}\right]\If(D_{it} = d)\If(u < Y_{it} < v)Y_{it} }{\sum_{i=1}^{N_t} \left[z\frac{Z_{it}}{e(X_{it})} + (1-z)\frac{1-Z_{it}}{1-e(X_{it})}\right]\If(D_{it} = d)}.
$$
In addition, $\hat c_{t,1}$ satisfies:
\begin{align*}
\tilde{F}_{t,Y_{it}(2)}(\hat c_{t,1}) = \frac{\hat \pi_t(0,2)}{\hat \pi_t(0,2)+ \hat \pi_t(1,2)} 
\end{align*}
where
\begin{align*}
\tilde{F}_{t,Y_{it}(2)}(\hat c_{t,1}) = & \frac{\hat \pi_t(0,2)+ \hat \pi_t(1,2) + \hat \pi_t(2,2)}{\hat \pi_t(0,2)+\hat \pi_t(1,2)}\hat F_{t,Y_{it}(2)}(\hat c_{t,1}|Z_{it}=1, D_{it}=2) \\
& -  \frac{\hat \pi_t(2,2)}{\hat \pi_t(0,2)+\hat \pi_t(1,2)}\hat F_{t,Y_{it}(2)}(\hat c_{t,1}|Z_{it}=0, D_{it}=2) \\
= & \frac{\hat \pi_t(0,2)+ \hat \pi_t(1,2) + \hat \pi_t(2,2)}{\hat \pi_t(0,2)+\hat \pi_t(1,2)}\frac{\sum_{i=1}^{N_t} \left[\frac{Z_{it}}{e(X_{it})}\right]\If(D_{it} = 2)\If(Y_{it} < \hat c_{t,1} )}{\sum_{i=1}^{N_t} \left[\frac{Z_{it}}{e(X_{it})} \right]\If(D_{it} = 2)} \\
& -  \frac{\hat \pi_t(2,2)}{\hat \pi_t(0,2)+\hat \pi_t(1,2)}\frac{\sum_{i=1}^{N_t} \left[\frac{1-Z_{it}}{1-e(X_{it})}\right]\If(D_{it} = 2)\If(Y_{it} < \hat c_{t,1} )}{\sum_{i=1}^{N_t} \left[\frac{1-Z_{it}}{1-e(X_{it})}\right]\If(D_{it} = 2)} \\
\end{align*}
The upper bound $\beta_{t,U}(0,2)$ can be similarly estimated and we omit the details to save space.

The conditional expectation $\mu^0_t(0,2)$ can be estimated via the following equation:

\begin{align*}
\hat \mu^0_t(0,2) = & \frac{\hat \pi_t(0,0)+\hat \pi_t(0,2)}{\hat \pi_t(0,2)}\frac{\sum_{i=1}^{N_t} \frac{1-Z_{it}}{1-e(X_{it})}\If(D_{it}=0)Y_{it}}{\sum_{i=1}^{N_t} \frac{1-Z_{it}}{1-e(X_{it})}\If(D_{it}=0)}\\
& -\frac{\hat \pi_t(0,0)}{\hat \pi_t(0,2)}\frac{\sum_{i=1}^{N_t} \frac{Z_{it}}{e(X_{it})} \If(D_{it}=0)Y_{it}}{\sum_{i=1}^{N_t} \frac{Z_{it}}{e(X_{it})} \If(D_{it}=0)}.
\end{align*}

Finally, we estimate the lower bound of the complier average causal effect, $\tau_t^L(0,2;(0,2))$, via $\hat \beta_{t,L}(0,2) - \hat \mu^0_t(0,2)$. Similarly, we can estimate the upper bound of $\tau_t(0,2;(0,2))$, as well as the bounds for the substitutor causal effect, $\tau_t(1,2;(1,2))$.

\section{Asymptotic inference for the bounds}
Now, we show the proposed estimators are consistent and asymptotically normal using the theory developed by.

Let's denote $\frac{\sum_{i=1}^{N_t} \frac{1-Z_{it}}{e(X_{it})}\If(D_{it} = 2)\If(Y_{it} < c_{t,1})}{\sum_{i=1}^{N_t} \frac{1-Z_{it}}{e(X_{it})}\If(D_{it} = 2)}$ as $\kappa_1$, $\frac{\sum_{i=1}^{N_t} \frac{Z_{it}}{e(X_{it})}\If(D_{it} = 0)Y_{it}}{\sum_{i=1}^{N_t} \frac{Z_{it}\If(D_{it} = 0)}{e(X_{it})}}$ as $\kappa_2$, and $\frac{\sum_{i=1}^{N_t} \frac{Z_{it}}{e(X_{it})}\If(D_{it} = 1)}{\sum_{i=1}^{N_t} \frac{Z_{it}}{e(X_{it})}}$ as $\kappa_3$ . We further denote parameters $\{\mu_t(-b_y, c_{t,1}; 1, 2), \mu_t(-b_y, c_{t,1}; 0, 2), \mu^0_t(0,2), c_{t,1}, \kappa_1, \kappa_2\}$ as $\theta$ and parameters $\{\pi_t(0,0), \pi_t(0,2), \pi_t(1,2), \pi_t(2,2), \kappa_3\}$ as $\gamma$. The parameter of interest, $\tau_t^L(0,2;(0,2))$, is equal to $\beta_{t,L}(0,2) - \mu^0_t(0,2) = \frac{\pi_t(0,2)+\pi_t(1,2)+ \pi_t(2,2)}{ \pi_t(0,2)}\mu_t(-b_y, c_{t,1}; 1, 2)  - \frac{\pi_t(2,2)}{\pi_t(0,2)} \mu_t(-b_y,  c_{t,1}; 0, 2) - \mu^0_t(0,2)$. The whole estimation process could be seen as a GMM estimator with the following moment conditions. For simplicity, we will ignore covariates.

\begin{align}
& Z_{it}\If(D_{it}=2)\left(\If (Y_{it} < c_{t,1})Y_{it} - \mu_t(-b_y, c_{t,1}; 1, 2)\right) \\
& (1-Z_{it})\If(D_{it}=2)\left(\If(Y_{it} < c_{t,1})Y_{it} - \mu_t(-b_y, c_{t,1}; 0, 2)\right) \\
& (1-Z_{it})\If(D_{it}=2)\left[\kappa_1 - \If(Y_{it} < c_{t,1})\right] \\
& Z_{it}\If(D_{it}=2)\left[\pi_t(2,2)\kappa_1 + \pi_t(0,2) - (\pi_t(0,2) + \pi_t(1,2) + \pi_t(2,2)) \If(Y_{it} < c_{t,1})\right] \\
& (1-Z_{it})\If(D_{it}=0)\left[(\pi_{t}(0,0) + \pi_{t}(0,2))Y_{it} - \pi_{t}(0,2)\mu^0_t(0,2) - \pi_{t}(0,0)\kappa_2 \right] \\ 
& Z_{it}\If(D_{it}=0)(Y_{it} - \kappa_2) \\
& Z_{it}\left[\pi_t(0,0) - \If(D_{it} = 0)\right] \\
& (1-Z_{it})\left[\pi_t(2,2) - \If(D_{it} = 2)\right] \\
& (1-Z_{it})\left[\pi_t(0,2) + \pi_t(0,0) - \If(D_{it} = 0)\right] \\
& (1-Z_{it})\left[\pi_t(1,2) + \kappa_3 - \If(D_{it} = 1)\right] \\
& Z_{it}\left[\kappa_3 - \If(D_{it} = 1)\right] 
\end{align}
Let $\mathbf{h} (w;\theta, \gamma) = (\mathbf{g} (w;\theta, \gamma), \mathbf{m} (w; \gamma))$ be the moment conditions above. We will rely on results in \citet{newey1994large} to show that the estimates are consistent and normally distributed as $N \rightarrow \infty$. 

For consistency, we apply Theorem 2.6 in \citet{newey1994large} to our $\mathbf{h} (w;\theta, \gamma)$. It is easy to easy that the equations lead to a unique solution. As long as $Y_{it}$ is bounded, we know that all the parameters, nuisance or not, are within a compact set. Hence, condition (ii) is satisfied. Condition (iii), the continuity of $\mathbf{h} (w;\theta, \gamma)$ is obvious and condition (iv) will follow.

Theorem 7.2 in \citet{newey1994large} is needed to show the asymptotic normality of the estimates. Condition (i) and (iii) are satisfied due to the same reason stated above. (iv) holds due to the central limit theorem. (ii) could be verified using matrix $\mathbf{H}$ described below. We use Theorem 1 in \citet{andrews1994asymptotics} to show the stochastic equicontinuity of $\mathbf{h} (w;\theta, \gamma)$. Assumption C is obvious. The envelop function is constructed following the example in \citet{lee2009training}. For moments (3) (4) (7)-(12), the envelop is set to be 3. For moments (1) and (2), the envelop is $|Y_{it} - \mu_{t0}| + \sup_{\mu_{t}} |\mu_{t0} - \mu_{t}|$ where $\mu_{t}$ is the corresponding parameter and $\mu_{t0}$ is its true value. For moment (5), the envelop is $|Y_{it}| + 1$. Then assumption A is satisfied. Assumption B holds as the outcome $Y_{it}$ is bounded.

Notice that $\mathbf{h}_0(w;\theta, \gamma) = \E\left[\mathbf{h} (w;\theta, \gamma)\right]$ equals to:
\begin{align*}
& P(D_{it}=2,  Z_{it}=1)\left( \int_{-b_y}^{c_{t,1}} y_{it} f_{Y_{it}|Z_{it=1}, D_{it}=2}(y_{it})d y_{it} - \mu_t(-b_y, c_{t,1}; 1, 2)\right) \\
& P(D_{it}=2, Z_{it}=0) \left(\int_{-b_y}^{c_{t,1}} y_{it} f_{Y_{it}|Z_{it=0}, D_{it}=2}(y_{it}) d y_{it} - \mu_t(-b_y, c_{t,1}; 0, 2)\right) \\
& P(D_{it}=2, Z_{it}=0)\left[\kappa_1 - F_{Y_{it} | D_{it}=2, Z_{it}=0}(c_{t,1})\right]   \\
& P(D_{it}=2, Z_{it}=1)\left[\pi_t(2,2)\kappa_1 + \pi_t(0,2) -  (\pi_t(0,2) + \pi_t(1,2) + \pi_t(2,2)) F_{Y_{it} | D_{it}=2, Z_{it}=1}(c_{t,1})\right]   \\
& P(D_{it}=0, Z_{it}=0)\left[(\pi_{t}(0,0) + \pi_{t}(0,2))\E(Y_{it}|D_{it}=0, Z_{it}=0) - \pi_{t}(0,2)\mu^0_t(0,2) - \pi_{t}(0,0)\kappa_2 \right] \\ 
& P(D_{it}=0, Z_{it}=1)\left[\E(Y_{it}|D_{it}=0, Z_{it}=1) - \kappa_2\right] \\
& P(Z_{it}=1)\left[\pi_t(0,0) - P(D_{it} = 0 | Z_{it}=1)\right] \\
& P(Z_{it}=0)\left[\pi_t(2,2) - P(D_{it} = 2 | Z_{it}=0)\right] \\
& P(Z_{it}=0)\left[\pi_t(0,2) + \pi_t(2,2) - P(D_{it} = 0 | Z_{it}=0)\right] \\
& P(Z_{it}=0)\left[\pi_t(1,2) + \kappa_3 - P(D_{it} = 1 | Z_{it}=0)\right] \\
& P(Z_{it}=1)\left[\kappa_3 - P(D_{it} = 1 | Z_{it}=1)\right] 
\end{align*}
We know that $\mathbf{h}_0(w;\theta_0, \gamma_0) = 0$, where $(\theta_0, \gamma_0)$ are the true values of $(\theta, \gamma)$. Define $\mathbf{H}$ as $\frac{d \mathbf{h}_0}{d (\theta, \gamma)'}|_{(\theta_0, \gamma_0)} = \begin{pmatrix}
\mathbf{G}_{\theta_0} & \mathbf{G}_{\gamma_0} \\
\mathbf{0} & \mathbf{M}_{\gamma_0}
\end{pmatrix}$. $\mathbf{H}$ is a $11 \times 11$ matrix. Similarly, define $\mathbf{\Sigma}$ as $\E \left[\mathbf{h} (w;\theta, \gamma) \mathbf{h}' (w;\theta, \gamma) \right]|_{(\theta_0, \gamma_0)} = \begin{pmatrix}
\mathbf{\Sigma}_{11} & \mathbf{\Sigma}_{12} \\
\mathbf{\Sigma}_{21} & \mathbf{\Sigma}_{22}
\end{pmatrix}$. Then, \citet{andrews1994asymptotics} indicates that the asymptotic variance is $\mathbf{V} = \mathbf{H}^{-1}\mathbf{\Sigma} (\mathbf{H}^{'})^{-1}$. As $\sqrt{N}\begin{pmatrix}
\hat \theta \\
\hat \gamma \\
\end{pmatrix} - \begin{pmatrix}
\theta_0 \\
\gamma_0
\end{pmatrix} \rightarrow N(0, \mathbf{V})$, it is easy to see that $\tau_t^L(0,2;(0,2))$ is also normally distributed whose variance can be caluclated via the delta method.

To derive the analytical variance, we treat the estimation as a two-step process. In the first step, all the $\pi$s are estimated. In the second step, the estimates are ``plugged in" and treated as known. Now the first six moments are used to estimate the parameters in the second step, and the last five moments are for the first-step parameters. To save space, we ignore the time index and write $\mu_t(-b_y, c_{t,1}; 1, 2)$ as $\mu_{12}$,  $\mu_t(-b_y, c_{t,1}; 0, 2)$ as $\mu_{02}$, $\mu^0_t(0,2)$ as $\mu^0$, $P(D_{it}=d)$ as $P_{Dd}$, $P(Z_{it}=z)$ as $P_{Zz}$, $P(D_{it}=d, Z_{it}=z)$ as $P_{DdZz}$, $F_{Y_{it}|Z_{it}=z, D_{it}=d}(c_{t,1})$ as $F_{dz}(c_1)$, $f_{Y_{it}|Z_{it}=z, D_{it}=d}(c_{t,1})$ as $f_{dz}(c_{1})$.

 Now the variance-covariance matrix for $\theta$ at $\theta_0$ is:
\begin{align*}
\mathbf{V}_{\theta_0} = \mathbf{G}_{\theta_0}^{-1}\E\left[\left(\mathbf{g} (w;\theta_0, \gamma_0)- \mathbf{G}_{\gamma_0}\mathbf{M}_{\gamma_0}^{-1}\mathbf{m} (w; \gamma_0)\right) \left(\mathbf{g} (w;\theta_0, \gamma_0)- \mathbf{G}_{\gamma_0}\mathbf{M}_{\gamma_0}^{-1}\mathbf{m} (w; \gamma_0) \right)^{'} \right] (\mathbf{G}_{\theta_0}^{'})^{-1}
\end{align*} 
It is easy to see:
\begin{align*}
\mathbf{G}_{\theta_0} = (\mathbf{G}_{1\theta_0}, \mathbf{G}_{2\theta_0})
\end{align*}
where

\begin{align*}
\mathbf{G}_{1\theta_0} = \begin{pmatrix}
-P_{D2Z1} & 0 & 0  \\
0 & -P_{D2Z0} & 0  \\
0 & 0 & 0  \\
0 & 0 & 0  \\
0 & 0 & -P_{D0Z0}\pi(0,2) \\
0 & 0 & 0 \\
\end{pmatrix}
\end{align*}
and

\begin{align*}
\mathbf{G}_{2\theta_0} = \begin{pmatrix}
P_{D2Z1}c_{1}f_{21}(c_{1}) & 0 & 0 \\
P_{D2Z0}c_{1} f_{20}(c_{1})  & 0 & 0 \\
-P_{D2Z0}f_{20}(c_{1}) & P_{D2Z0} & 0 \\
 -P_{D2Z1} (\pi(0,2) + \pi(1,2) + \pi(2,2))f_{21}(c_{1}) & P_{D2Z1}\pi(2,2) & 0 \\
0 & 0 & -P_{D0Z0}\pi(0,0) \\
0 & 0 & -P_{D0Z1} \\
\end{pmatrix}
\end{align*}

Let $a_1 = - (\pi(0,2) + \pi(1,2) + \pi(2,2))f_{21}(c_{t,1}) + \pi(2,2)f_{20}(c_{1})$, then we can write the inverse of $G_{\theta_0}$ as:
\begin{align*}
&\mathbf{G}_{\theta_0}^{-1} = (\mathbf{G}_{1\theta_0}^{-1}, \mathbf{G}_{2\theta_0}^{-1})
\end{align*}
where

\begin{align*}
\mathbf{G}_{1\theta_0}^{-1} = \begin{pmatrix}
\frac{1}{-P_{D2Z1}} & 0 & -\frac{\pi(2,2)c_{1} f_{21}(c_{1})}{a_1 P_{D2Z0}} \\
0 & -\frac{1}{P_{D2Z0}} & -\frac{\pi(2,2)c_{1} f_{20}(c_{1})}{a_1 P_{D2Z0}}  \\
0 & 0 & 0  \\
0 & 0 & -\frac{\pi(2,2)}{a_1 P_{D2Z0}} \\
0 & 0 & \frac{1}{P_{D2Z0}} - \frac{\pi(2,2)f_{20}(c_{t,1})}{a_1 P_{D2Z0}}  \\
0 & 0 & 0 
\end{pmatrix}
\end{align*}
and

\begin{align*}
\mathbf{G}_{2\theta_0}^{-1} = \begin{pmatrix}
\frac{c_{1} f_{21}(c_{1})}{a_1 P_{D2Z1}} & 0 & 0 \\
\frac{c_{1} f_{20}(c_{1})}{a_1 P_{D2Z1}} & 0 & 0  \\
0 & -\frac{1}{P_{D0Z0}\pi(0,2)} & \frac{\pi(0,0)}{P_{D0Z1}\pi(0,2)}  \\
\frac{1}{a_1 P_{D2Z1}} & 0 & 0  \\
\frac{f_{20}(c_{1})}{a_1 P_{D2Z1}} & 0 & 0 \\
0 & 0 & -\frac{1}{P_{D0Z1}}  \\
\end{pmatrix}
\end{align*}
Similarly, $\mathbf{\Sigma}_{11} = \E \left[\mathbf{g} (w;\theta_0, \gamma_0) \mathbf{g}' (w;\theta_0, \gamma_0) \right] = (\mathbf{\Sigma}_{11,1}, \mathbf{\Sigma}_{11,2})$. Let $a_2 = \pi(2,2)\kappa_1 + \pi(0,2)$, $a_3 = \pi(0,2) + \pi(1,2) + \pi(2,2)$, $a_4 = \pi(0, 0) + \pi(0, 2)$, $a_5 = \pi(0,2)\mu^0 + \pi(0,0)\kappa_2$, $\E_{dz}\left[Y_i^k\right] = \E\left[Y_i^k | D_i = d, Z_i = z \right]$, and $\Var_{dz}\left[Y_i^k\right] = \Var\left[Y_i^k | D_i = d, Z_i = z \right]$, then:
\begin{scriptsize}
\begin{align*}
& \mathbf{\Sigma}_{11,1} = \\
& \begin{pmatrix}
P_{D2Z1}\left( \int_{-b_y}^{c_{1}} y_{i}^2  f_{21}(y_{i})d y_{i}- \mu_{12}^2\right) & 0 & 0 \\
0 & P_{D2Z0}\left(\int_{-b_y}^{c_{1}} y_{t}^2 f_{20}(y_{i})d y_{i}- \mu_{02}^2\right) & (\kappa_1 - 1) P_{D2Z0}\left( \int_{-b_y}^{c_{1}} y_{i} f_{20}(y_{i})d y_{i} - \mu_{02}\right) \\
0 & (\kappa_1 - 1) P_{D2Z0}\left(\int_{-b_y}^{c_{1}} y_{i} f_{20}(y_{i})d y_{i} - \mu_{02}\right) & P_{D2Z0}(\kappa_1 -  F_{20}(c_1))^{2}  \\
(a_2 - a_3) P_{D2Z1} \left(\int_{-b_y}^{c_{1}} y_{i} f_{21}(y_{i})d y_{i} - \mu_{12}\right) & 0 & 0  \\
0 & 0 & 0 \\
0 & 0 & 0  \\
\end{pmatrix}
\end{align*}
\end{scriptsize}
and
\begin{scriptsize}
\begin{align*}
& \mathbf{\Sigma}_{11,2} = \\
& \begin{pmatrix}
(a_2 - a_3) P_{D2Z1}\left(\int_{-b_y}^{c_{1}} y_{i} f_{21}(y_{i})d y_{i} - \mu_{12}\right) & 0 & 0 \\
0 & 0 & 0 \\
0 & 0 & 0  \\
P_{D2Z1}(a_2^2 - 2 a_2 a_3F_{21}(c_1) + a_3^2 F_{21}(c_1))  & 0 & 0  \\
0 & P_{D0Z0}\E_{00}\left[a_4 Y_i - a_5 \right]^2 & 0 \\
0 & 0  & P_{D0Z1}\E_{01}\left[Y_i - \kappa_4\right]^2  \\
\end{pmatrix}
\end{align*}
\end{scriptsize}
Exploiting the fact that moment conditions as well as the fact that their expectations equal to zero, we have:
\begin{scriptsize}
\begin{align*}
& \mathbf{\Sigma}_{11,1} = \\
& \begin{pmatrix}
P_{D2Z1}\left( \int_{-b_y}^{c_{1}} y_{i}^2  f_{21}(y_{i})d y_{i}- \mu_{12}^2\right) & 0 & 0 \\
0 & P_{D2Z0}\left(\int_{-b_y}^{c_{1}} y_{t}^2 f_{20}(y_{i})d y_{i}- \mu_{02}^2\right) & 0 \\
0 & 0 & P_{D2Z0}\kappa_1(1-\kappa_1)  \\
0 & 0 & 0  \\
0 & 0 & 0  \\
0 & 0 & 0 
\end{pmatrix}
\end{align*}
\end{scriptsize}
and
\begin{scriptsize}
\begin{align*}
& \mathbf{\Sigma}_{11,2} = \\
& \begin{pmatrix}
0 & 0 & 0  \\
0 & 0 & 0  \\
0 & 0 & 0  \\
P_{D2Z1}a_2 (a_3 - a_2) & 0 & 0 \\
0 & P_{D0Z0}a_4^2 \Var_{00}\left[Y_i \right] & 0 \\
0 & 0 & P_{D0Z1}\Var_{01}\left[Y_i \right]
\end{pmatrix}
\end{align*}
\end{scriptsize}
Now, $\mathbf{\tilde{V}}_{\theta_0} = \mathbf{G}_{\theta_0}^{-1}\mathbf{\Sigma}_{11} (\mathbf{G}_{\theta_0}^{'})^{-1}$ is the variance of $\theta$ at $\theta_0$
if we ignore the uncertainty induced by the first stage. In this case, we have:

\begin{align*}
& \Var(\mu_t(-b_y, c_{t,1}; 1, 2)) = \frac{\int_{-b_y}^{c_{1}} y_{i}^2 f_{21}(y_{i})d y_{i}  - \mu_{12}^2}{P_{D2Z1}} \\
& \Var(\mu_t(-b_y, c_{t,1}; 0, 2)) = \frac{\int_{-b_y}^{c_{1}} y_{i}^2 f_{20}(y_{i})d y_{i}  - \mu_{02}^2}{P_{D2Z0}} \\
& \Cov(\mu_t(-b_y, c_{t,1}; 1, 2), \mu_t(-b_y, c_{t,1}; 0, 2)) = 0
\end{align*}
and
\begin{align*}
\Var( \mu^0_t(0,2)) = & \frac{(\pi(0,0) + \pi(0,2))^2}{P_{D0Z0}\pi^2(0,2)} \Var_{00}\left[Y_i\right] \\
& + \frac{\pi^2(0,0)}{P_{D0Z1}\pi^2(0,2)} \Var_{01}\left[Y_i\right] \\
\end{align*}
\begin{align*}
& \Cov(\mu_t(-b_y, c_{t,1}; 1, 2), \mu^0_t(0,2)) = 0 \\
& \Cov(\mu_t(-b_y, c_{t,1}; 0, 2), \mu^0_t(0,2)) = 0
\end{align*}

To proceed, we have:
\begin{footnotesize}
\begin{align*}
& \mathbf{G}_{\gamma} = \\
& \begin{pmatrix}
0 & 0 & 0 & 0 & 0  \\
0 & 0 & 0  & 0 & 0  \\
0 & 0 & 0  & 0 & 0  \\
0 & P_{D2Z1}(1-F_{21}(c_1)) & -P_{D2Z1}F_{21}(c_1) & P_{D2Z1}\left(\kappa_1 - F_{21}(c_1) \right) & 0 \\
P_{D0Z0}(\E_{00}Y_i - \kappa_4) & P_{D0Z0}(\E_{00}Y_i - \mu^0) & 0 & 0 & 0 \\
0 & 0 & 0  & 0 & 0 \\
\end{pmatrix}
\end{align*}
\end{footnotesize}

\begin{align*}
\mathbf{M}_{\gamma} = \begin{pmatrix}
P_{Z1}& 0 & 0 & 0 & 0 \\
0 & 0 & 0  & P_{Z0} & 0 \\
0 & P_{Z0} & 0 & P_{Z0} & 0 \\
0 & 0 & P_{Z0} & 0 & P_{Z0} \\
0 & 0 & 0 & 0 & P_{Z1}
\end{pmatrix}
\end{align*}
and

\begin{scriptsize}
\begin{align*}
& \mathbf{\Sigma}_{22} = \E\left[\mathbf{m} (w; \gamma) \mathbf{m}^{'} (w; \gamma)\right] \\
& = \begin{pmatrix}
P_{Z1}(\pi_{00} - \pi_{00}^2) & 0 & 0 & 0 & -P_{Z1}\kappa_3 \pi_{00} \\
0 & P_{Z0}(\pi_{22} - \pi_{22}^2) & -P_{Z0}\pi_{22}(\pi_{00} + \pi_{02}) & -P_{Z0}\pi_{22}(\pi_{12}+\kappa_3) & 0  \\
0 & -P_{Z0}\pi_{22}(\pi_{00} + \pi_{02}) & P_{Z0}(p_{00} - p_{00}^2) & -P_{Z0}\pi_{12}(\pi_{00} + \pi_{02}) & 0 \\
0 & -P_{Z0}\pi_{22}(\pi_{12}+\kappa_3) & -P_{Z0}\pi_{12}(\pi_{00} + \pi_{02}) & P_{Z0}(p_{10} - p_{10}^2) & 0 \\
-P_{Z1}\kappa_3 \pi_{00} & 0 & 0 &  0 & P_{Z1}(p_{11} - p_{11}^2) 
\end{pmatrix}
\end{align*}
\end{scriptsize}
We also know that $\mathbf{\Sigma}_{12} = \mathbf{\Sigma}_{21} = \mathbf{0}$. Then the variance-covariance matrix $\mathbf{V}$ can be calculated as $\mathbf{H}^{-1}\mathbf{\Sigma} (\mathbf{H}^{'})^{-1}$.

Finally, as $\tau_t^L(0,2;(0,2)) = f(\mu_t(-b_y, c_{t,1}; 1, 2), \mu_t(-b_y, c_{t,1}; 0, 2), \mu^0_t(0,2), \pi_t(0,2), \pi_t(1,2), \pi_t(2,2)) = f(\eta)$, we can use the delta method to derive the asymptotic distribution of $\tau_t^L(0,2;(0,2))$:
\begin{align*}
\Var(\tau_t^L(0,2;(0,2))) = (\frac{d f}{d \eta})^{'} \mathbf{V}_{\eta} \frac{d f}{d \eta}
\end{align*}
where $\frac{d f}{d \eta} = (\frac{\pi_{02} + \pi_{12} + \pi_{22}}{\pi_{02}}, -\frac{\pi_{22}}{\pi_{02}}, -1, -\frac{\pi_{12} + \pi_{22}}{\pi_{02}^2}\mu_{12} +\frac{\pi_{22}}{\pi^2_{02}}\mu_{02}, \frac{1}{\pi_{02}}\mu_{12}, \frac{1}{\pi_{02}}(\mu_{12} - \mu_{02}))^{'}$,
and $\mathbf{V}_{\eta}$ is the corresponding submatrix of $\mathbf{V}$. The variances of $\tau_t^U(0,2;(0,2))$, $\tau_t^L(1,2;(1,2))$, and $\tau_t^U(1,2;(1,2))$ can be similarly obtained.

\end{appendix}

\clearpage
\bibliography{alse}

\end{document}